\documentclass[conference]{IEEEtran}
\IEEEoverridecommandlockouts
\IEEEoverridecommandlockouts
\usepackage{cite}
\usepackage{amsmath,amssymb,amsfonts}
\usepackage{algorithm}  

\usepackage{graphicx}
\usepackage{textcomp}
\usepackage{xcolor}
\usepackage{algorithm}
\usepackage{algorithmic}
\usepackage{balance}
\usepackage{longtable}
\usepackage{array}
\usepackage{float}
\usepackage{textcomp}
\usepackage{bbding}
\usepackage{url}
\usepackage[colorlinks=true,allcolors=blue,bookmarks=false]{hyperref}
\def\BibTeX{{\rm B\kern-.05em{\sc i\kern-.025em b}\kern-.08em
    T\kern-.1667em\lower.7ex\hbox{E}\kern-.125emX}}
    
\begin{document}

\title{Serverless Edge Computing: A Taxonomy, Systematic Literature Review, Current Trends and Research Challenges\\}




\author{\IEEEauthorblockN{Iqra Batool}
\IEEEauthorblockA{\textit{Department of Computer Science} \\
\textit{University of Western Ontario}\\
London, ON, Canada \\
ibatool2@uwo.ca}
\and
\IEEEauthorblockN{Sania Kanwal}
\IEEEauthorblockA{\textit{Information Engineering} \\
\textit{Southwest University of Science and Technology}\\
Mianyang 621010, China\\
saniakanwal398@yahoo.com}
}

\maketitle
\begin{abstract}
In recent years, the rapid expansion of Internet of Things (IoT) nodes and devices has seamlessly integrated technology into everyday life, amplifying the demand for optimized computing solutions. To meet the critical Quality of Service (QoS) requirements—such as reduced latency, efficient bandwidth usage, swift reaction times, scalability, privacy, and security—serverless edge computing has emerged as a transformative paradigm. This systematic literature review explores the current landscape of serverless edge computing, analyzing recent studies to uncover the present state of this technology. The review identifies the essential features of serverless edge computing, focusing on architectural designs, QoS metrics, implementation specifics, practical applications, and communication modalities central to this paradigm. Furthermore, we propose a comprehensive taxonomy that categorizes existing research efforts, providing a comparative analysis based on these classifications. The paper concludes with an in-depth discussion of open research challenges and highlights promising future directions that hold potential for advancing serverless edge computing research.
\end{abstract} 

\maketitle

\ifCLASSOPTIONcompsoc
\IEEEraisesectionheading{\section{Introduction}\label{sec:introduction}}
\else
\section{Introduction}
\label{sec:introduction}
With the continuous emergence of mobile applications like the Internet of Things (IoT), augmented reality (AR), and self-driving cars, the volume of mobile data generated from multiple distributed sources is growing at an exponential rate~\cite{tang2020decentralized-1}. To manage this surge in data, cloud computing has surfaced as a promising solution, offering advantages such as on-demand self-service, resource sharing, and dynamic scalability~\cite{varghese2018next-2}. However, because cloud servers are often located far from end users, the latency incurred from sending and retrieving large data payloads can become prohibitively high. To address these challenges, edge computing has emerged as a means to reduce both latency and performance bottlenecks by bringing processing closer to data sources~\cite{mao2017survey-3}.

Building upon these advancements, serverless computing has introduced a powerful approach to addressing issues related to scalability, load balancing, and resource allocation~\cite{chaudhry2020improved-4}. Since the release of AWS Lambda, several major cloud providers have extended serverless computing capabilities to the edge, with platforms such as Apache OpenWhisk, Google Cloud Functions, and OpenFaaS among others~\cite{cicconetti2020decentralized-5}, ~\cite{ahmad2021container-7}. By offloading operational tasks—such as provisioning and scaling—to the service provider, serverless computing allows developers to focus on function implementation without managing underlying resources.

Research in serverless edge computing continues to advance, with a wealth of tutorials and research articles published in recent years. Various architectures and frameworks have been introduced, specifically tailored to support diverse IoT applications~\cite{aslanpour2021serverless-21},\cite{glikson2017deviceless-10}. Despite this progress, the current literature lacks a cohesive taxonomy for serverless edge computing, making it challenging to fully grasp the technology's state and identify existing research gaps. Thus, developing a taxonomy for serverless edge computing is essential to provide a comprehensive overview of this field and to highlight areas for future exploration. This paper presents an in-depth literature review of serverless edge computing architectures and frameworks, offering a foundation for researchers to better understand the field and to guide their continued work in serverless edge computing.
\subsection{Our Contribution}
In this study, we conduct a systematic literature review (SLR) on serverless edge computing, adhering closely to the guidelines established by Kitchenham~\cite{kitchenham2004procedures-11}. The primary goal of this research is to explore and analyze key attributes of serverless edge computing to uncover the specific challenges related to its architectural design, Quality of Service (QoS) metrics, application domains, implementation strategies, and communication methods. Building upon insights from the literature, we propose a comprehensive taxonomy for serverless edge computing, offering a structured categorization of previous research efforts and enabling a clearer comparison across studies within the field. Additionally, this review presents a synthesis of the advancements needed to drive further development and implementation of effective architectures and frameworks for serverless edge computing. The key contributions of our research can be summarized as follows:
\begin{itemize}
    \item Perform a systematic literature review on serverless edge computing.
    \item Describe the current status of search in serverless edge computing.
    \item Proposed a detailed taxonomy for classifying serverless edge computing  study into several categories.
    \item Compare prior research based on various factors related to the specified categories.
    \item Outlined the key research issues and possible future research directions.
\end{itemize}

Figure~\ref{fig: Structure} illustrates the overall structure of our review article. The remainder of the paper is structured as follows. Section~\ref{background} discusses the Serverless edge computing background information, and it is further divided into three sections which elaborate on the context, motivation, and related work. Section~\ref{ReviewMethodology} explains the review methodology used for this systematic literature review. The current trend of serverless edge computing, analysis of primary studies, and results outcomes are discussed in section~\ref{Current State}. In section~\ref{results and discussion}, we discussed the results, and section~\ref{Current Trend} presented the primary research challenges and future directions. Finally, section~\ref{Conclusion} concludes the review paper.
\begin{figure*}[h]
    \centering
    \includegraphics[width=\textwidth]{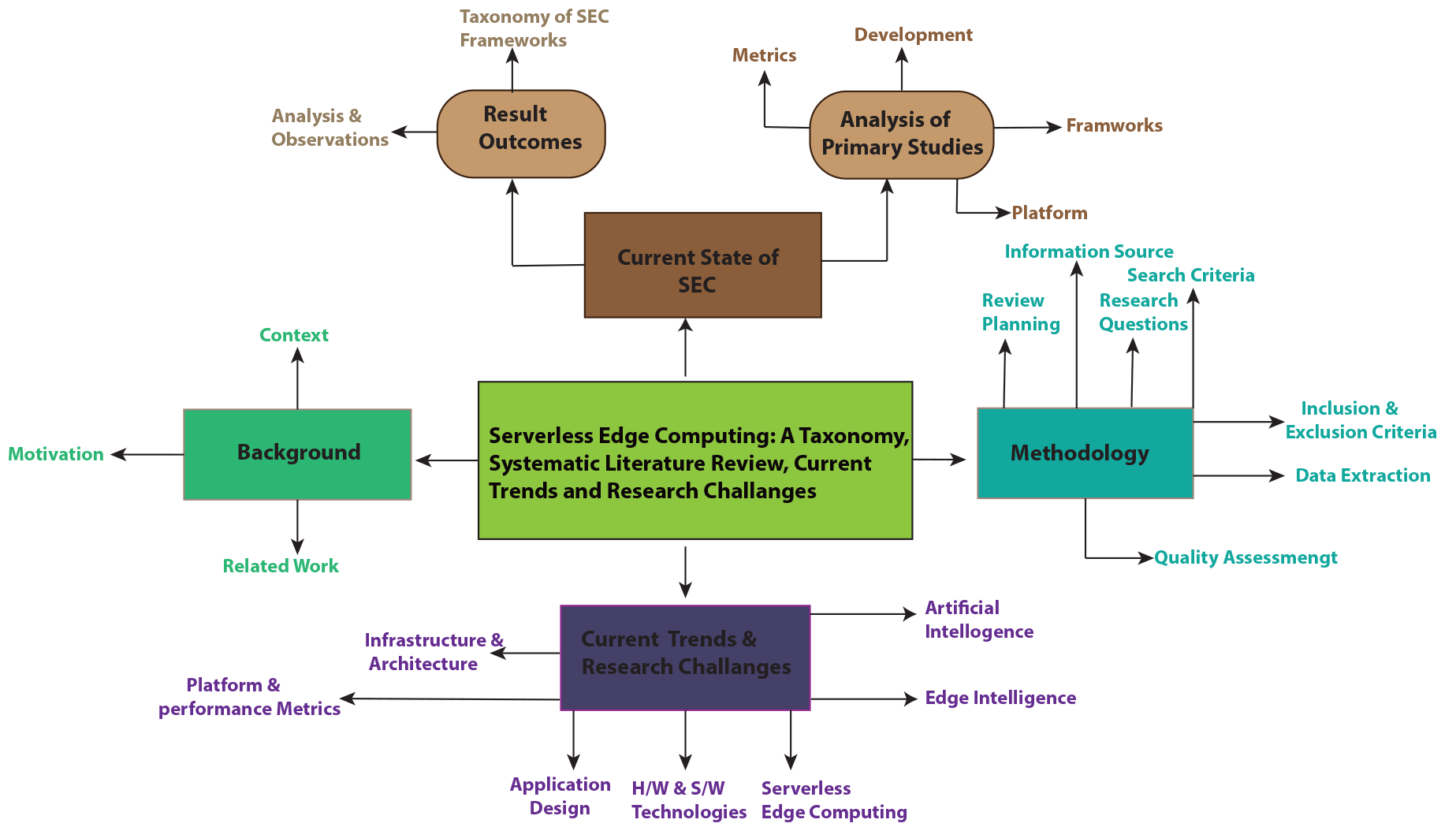}
    \caption{The organization of related work}
    \label{fig: Structure}
\end{figure*}

\section{Background}\label{background}
This section provides the context of serverless edge computing, motivation, and related work.
\subsection{Context}
The rapid growth of the Internet of Things (IoT) devices, mobile internet, and interconnected gadgets is generating vast amounts of data, with estimates suggesting that by the next few years, between sixty to seventy billion devices will be internet-enabled, amplifying data volumes considerably~\cite{dabbaghjamanesh2020real-12}. Managing this surge with conventional computing paradigms, such as cloud and distributed computing, is challenging due to limitations in handling ultra-high latency, bandwidth constraints, and the geographic dispersion of data~\cite{gill2019transformative-13}. Moreover, real-time response is essential in applications like emergency response in smart healthcare, traffic systems, and smart grids, where delays could compromise functionality and mobility support~\cite{tuli2020healthfog-14}. Hence, a complementary computing paradigm is needed that can process IoT device requests with minimal latency and response time to address these constraints effectively~\cite{xie2021serverless-15}.

The introduction of serverless computing, led by Amazon's launch of AWS Lambda in 2014, has opened up new opportunities for flexible, event-driven computing. This trend has been embraced by major IT providers, with Microsoft Azure Functions, Google Cloud Functions, Oracle Functions, and IBM Cloud Functions expanding the serverless landscape to meet the needs of an evolving marketplace. The research community has responded by developing user-friendly platforms like OpenFaaS, Fission, Kubeless, and IronFunctions, advancing serverless technology and its accessibility.

Serverless edge computing, in particular, is gaining traction across various industries as a promising approach to deploying and managing Function-as-a-Service (FaaS) solutions at the edge, addressing the unique challenges of edge nodes such as limited resources and complex management needs~\cite{mohanty2018evaluation-16}. Frameworks like SAM, Chalice, and Serverless.com have emerged to streamline serverless deployments. Open-source projects are advancing flexibility and configuration ease, especially on edge devices. For instance, OpenFaaS has developed a lightweight version called faasd, optimized for single-board computers (SBCs) like Raspberry Pi, which facilitates efficient edge implementations. Figure~\ref{fig: SEC ARc} illustrates the architecture of serverless edge computing.
\begin{figure}[h]
    \centering
    \includegraphics[width=9cm]{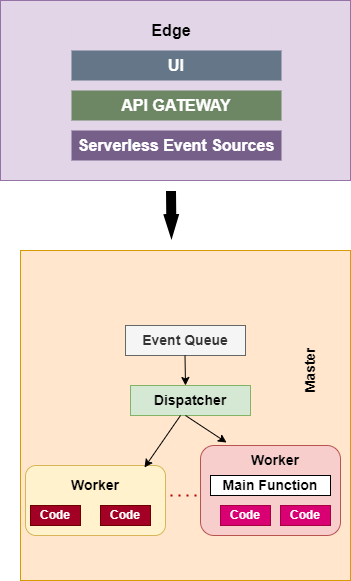}
    \caption{Serverless Edge Computing Architecture}
    \label{fig: SEC ARc}
\end{figure}
\subsection{Motivation}
As Internet of Things (IoT) applications proliferate, the need for efficient support of these applications has intensified. Essential requirements include low latency, real-time execution, event-driven programming, and rapid deployment, though additional demands continue to emerge~\cite{aslanpour2020performance-17}. Serverless edge computing has become instrumental in meeting these needs by enabling low-latency interactions between devices, thereby reducing response time and accelerating data processing.

Applications hosted on serverless computing platforms are constructed from small, independent code segments known as "functions." These functions, typically stateless, encapsulate the core application logic~\cite{mampage2022holistic-18}. An application may consist of a single function or several interconnected functions, depending on its complexity. Users define the workflow dependencies and abstract resource requirements~\cite{jonas2019cloud-19}, while the provider manages runtime orchestration to ensure efficient execution. Serverless platforms leverage distributed resources across multiple infrastructure domains, including public clouds, on-premises private clouds, and emerging architectures like fog and edge computing. Edge computing resources, such as sensors, smartphones, and smart devices, contribute their spare computing and storage capacity. Meanwhile, fog computing resources, which operate between the edge and cloud, offer intermediate levels of computing, storage, and memory capacity.

For serverless platforms in production, resource management decisions must balance the needs of both users and providers. Quality of Service (QoS) objectives, such as performance, cost-efficiency, and security, are critical for users. Providers, on the other hand, aim for high throughput by optimizing resource utilization. However, increasing resource utilization can sometimes impact QoS, creating a competitive tension between these goals. Balancing these priorities is crucial to achieving an efficient and reliable serverless system that meets both user and provider expectations.
\subsection{Related Work}\label{Relatedwork}
In this section, we review existing studies on serverless edge computing frameworks and architectures. To the best of our knowledge, there is no comprehensive systematic literature review on the taxonomy of serverless edge computing. Various authors have conducted literature reviews on serverless computing in specific IoT application areas~\cite{cassel2022serverless-20}. Notably, some surveys focus on serverless edge computing~\cite{aslanpour2020performance-17},\cite{ioini2020platforms-22},\cite{xie2021serverless-15}, while others explore serverless computing within cloud environments~\cite{taibi2020patterns-24},\cite{eismann2020review-25},\cite{yussupov2019systematic-26},\cite{scheuner2020function-27},\cite{li2022serverless-38},\cite{mampage2022holistic-18}. These surveys on serverless edge computing, however, lack rigorous methodologies, as they do not specify the research strategies used to select the evaluated papers. Therefore, they are best classified as literature reviews rather than systematic reviews.

~\cite{aslanpour2021serverless-21} provides valuable insights into serverless edge computing, demonstrating how edge devices benefit from this technology. The authors discuss opportunities and open issues, including task offloading, location-agnostic processing, energy efficiency, and security. Similarly,~\cite{ioini2020platforms-22} presents a review of platforms suitable for serverless edge computing, highlighting core features and suggesting avenues for future research. Platforms such as AWS IoT Greengrass, Azure IoT Edge, FogFlow, Nuclio, and OpenWhisk-Light are identified as supportive of edge deployments.

In another survey,~\cite{xie2021serverless-15} provides an in-depth analysis of serverless edge computing architectures, challenges, and unresolved issues. While these surveys address various aspects of serverless edge computing, others focus on traditional serverless computing on cloud platforms, discussing issues and solutions relevant to cloud-only environments.

Further contributions to this area include~\cite{kjorveziroski2021iot-48}, who perform a systematic literature review of IoT serverless computing at the edge, examining eight areas, including efficiency, application implementation, scheduling, benchmarks, platform implementation, policy, and open-source software. This study highlights the growing interest in serverless edge computing over the last three years. Similarly,~\cite{gadepalli2019challenges-49} discusses open challenges in serverless edge computing, such as existing serverless designs' limitations. The study explores WebAssembly as an alternative to traditional implementations, offering near-native performance, low memory use, and instant invocation in serverless applications.

Meanwhile,~\cite{li2022serverless-38} surveys serverless design architectures, categorizing them into four layers—virtualization, encapsulation, system orchestration, and coordination. This layered approach examines the benefits and limitations of each layer, proposing solutions to current challenges. Complementing this,~\cite{mampage2022holistic-18} reviews resource management in serverless computing, proposing a taxonomy that encompasses resource management across edge, fog, and cloud infrastructures, focusing on system design, workload management methodologies, and QoS objectives.

In their "multivocal literature review,"~\cite{taibi2020patterns-24} identifies 32 patterns in serverless solution deployment, grouped into orchestration, event management, availability, communication, and permission. They discuss the advantages and challenges of each pattern. Additionally,~\cite{eismann2020review-25} presents attributes from 89 real-world serverless use cases, noting that most solutions rely on AWS, consist of fewer than five functions, and are commonly implemented in JavaScript or Python. Interestingly, only 3\% of these cases address serverless edge computing, revealing a significant research gap in this domain.

~\cite{raith2023serverless} provides a structured review of serverless edge computing frameworks, focusing on the edge–cloud continuum's unique challenges, including programming support for AI, reliability, and performance optimization. The authors assess the maturity of various commercial, open-source, and academic platforms using a set of defined criteria that evaluate aspects such as application state management, fault tolerance, and resource efficiency. Unlike previous studies, this work introduces a maturity model to systematically analyze platform capabilities, emphasizing the importance of AI-driven intelligence, SLO awareness, and edge support. Their findings highlight gaps in existing frameworks and offer a roadmap for advancing serverless edge computing, aiming to unify cloud and edge resources effectively and enable robust, latency-sensitive IoT applications.
Table~\ref{tab:RelatedWork} shows the related work
\begin{center}
\begin{table*}
\caption{Related Work}
\label{tab:RelatedWork}
\begin{tabular}{|p{1.8cm}|c|c|c|p{1.8cm}|p{7.8cm}|}
\hline
\textbf{Authors} & \textbf{SLR} & \textbf{LR} & \textbf{Tax.} & \textbf{Main Focus} & \textbf{Description} \\
\hline
\cite{aslanpour2021serverless-8} & -- & \checkmark & \checkmark & Edge & Opportunities and challenges for serverless edge computing. \\
\hline
\cite{ioini2020platforms-22} & -- & \checkmark & -- & Edge & Discuss serverless edge computing platforms. \\
\hline
\cite{xie2021serverless-15} & -- & \checkmark & -- & Edge & Discuss opportunities and open challenges for serverless edge computing. \\
\hline
\cite{li2022serverless-38} & -- & \checkmark & -- & Serverless & Survey of serverless computing research domains, highlighting issues and future work suggestions. \\
\hline
\cite{kjorveziroski2021iot-48} & \checkmark & -- & -- & Serverless & Discussion of open issues for IoT serverless at the edge. \\
\hline
\cite{gadepalli2019challenges-49} & -- & \checkmark & -- & Serverless & Analysis of open issues and benefits of serverless at the edge. \\
\hline
\cite{cassel2022serverless-20} & \checkmark & -- & \checkmark & Cloud & Systematic review of serverless computing for IoT, covering characteristics and architecture of serverless cloud computing. \\
\hline
\cite{taibi2020patterns-24} & \checkmark & -- & \checkmark & Cloud & Organization patterns for serverless solutions. \\
\hline
\cite{eismann2020review-25} & \checkmark & -- & -- & Cloud & Serverless computing with real-world use cases. \\
\hline
\cite{yussupov2019systematic-26} & \checkmark & -- & -- & Cloud & Current trends, challenges and industry relationships in serverless computing.\\
\hline
\cite{scheuner2020function-27} & \checkmark & -- & \checkmark & Cloud & Performance evaluation of FaaS platforms. \\
\hline
\end{tabular}
\end{table*}
\end{center}

The authors~\cite{yussupov2019systematic-26} propose a Systematic Mapping Study that covers 62 papers from 2009 to 2019 and focuses on the following three primary subjects: first, publication trends regarding improvements or proposals of FaaS platforms and tools; second, classification of papers based on the challenges, and finally discovering how the industry is connected to the publications, as some papers focus on prototypes while others present production-ready solutions that can be effectively used on the industry. Finally,~\cite {scheuner2020function-27} presented a literature review that contains 112 distinct research that addressed performance measures for different serverless platforms.
 According to an  existing survey, the key shortcomings are  as follows
 \begin{enumerate}
     \item  According to the current studies, they did not provide any taxonomy or systematic literature for serverless edge computing.
     \item The current surveys have not included important factors for serverless edge computing framework.
     \item The surveys that were conducted previously were done so without using or just partly using a systematic literature review process. In this study, we adhered to the protocol for systematic reviews and used genuine articles as our criterion for article selection.
 \end{enumerate}
 \section{Review Methodology}\label{ReviewMethodology}
 This article presents a systematic literature review on serverless edge computing according to the recommendations of Kitchenham et al.~\cite{brereton2007lessons-28},\cite{kitchenham2009systematic-29},\cite{kitchenham2013systematic-30}. The systematic review process of this paper includes the following steps.
 \begin{enumerate}
     \item Defining the procedure for review
     \item Defining the rules for review
     \item Taxonomy design
     \item Performing a review
     \item Comparison with several existing techniques
     \item Discuss the result outcomes
     \item Finally, highlighting future research directions
 \end{enumerate}
 \subsection{Review Planning}\label{Review Planning}
 To develop norms for the review, first, research questions are defined. Then, we conducted more searches on these formulated queries in several databases. The review method is used to identify and gather adequate information for the desired research. The review process plays a significant role in determining which articles are taken into consideration or not. In this situation, the inculcation of bias might happen by choice of a single researcher in this work. Research questions for the review are shown in Table~\ref{tab: Req table}. Different search strings are used for the review process, shown in Table~\ref{tab: Search String}. The review process is shown in Figure~\ref{fig: keyword image}
 \begin{figure}[h]
     \centering
     \includegraphics[width = 11cm]{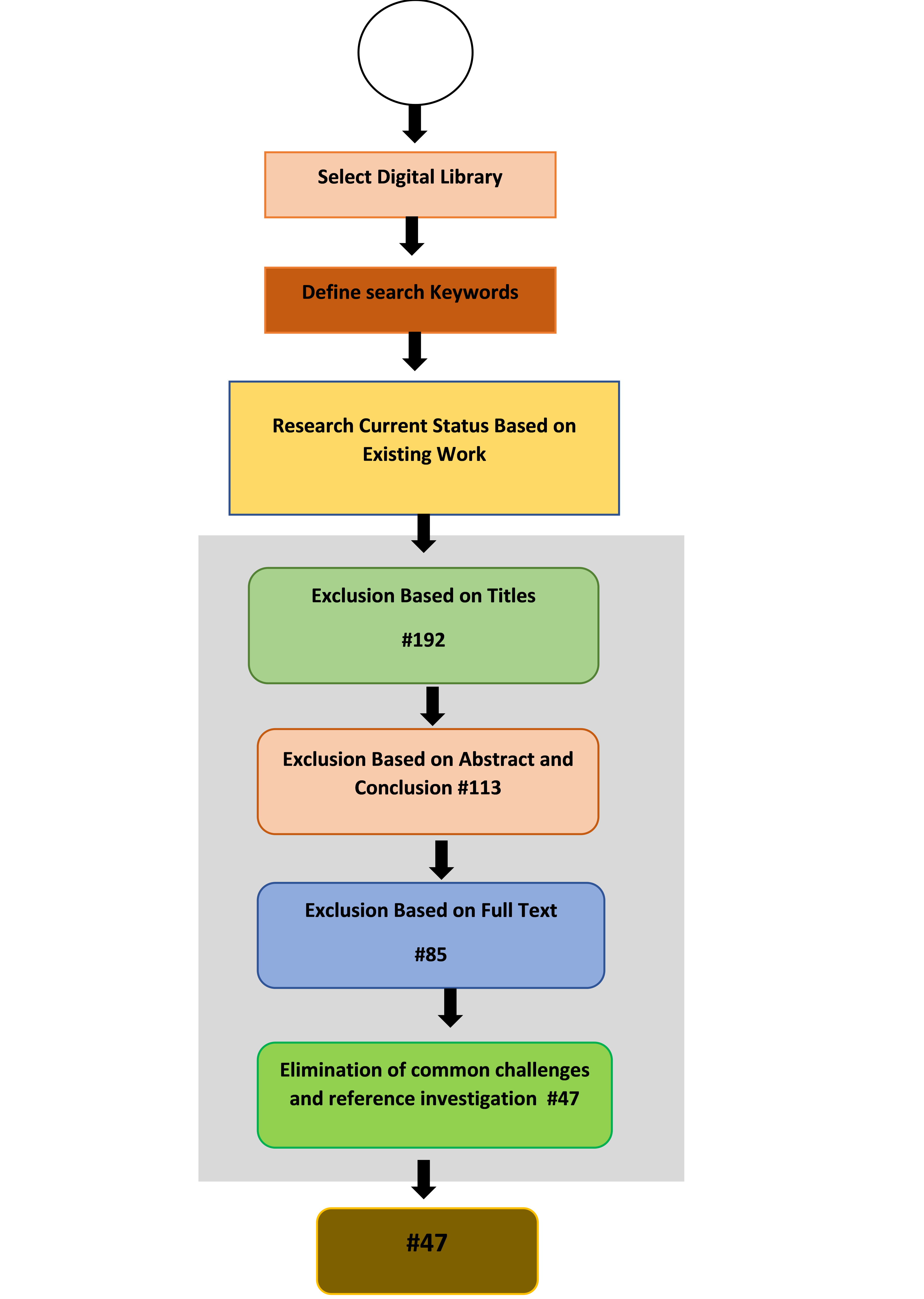}
     \caption{ Selection and Review Process for Primary Studies}
     \label{fig: keyword image}
 \end{figure}
 \subsection{Research Questions}\label{Research Questions}
 The objective of our study is to perform a systematic literature review on serverless edge computing. This study will help the researchers to understand the current state of serverless edge computing and outline possible future research directions. For the planning of the review process, research questions are required. The research questions, motivation, and correlations are shown in Table~\ref{tab: Req table}, demonstrating how our systematic literature review addresses these research questions.
\begin{center}
\begin{table*}[]
\caption{Research questions, Objective, Categorization, and relevant section }
\label{tab: Req table}
\begin{tabular}{|p{1.9 cm}|p{3.5 cm}|p{5.5 cm }|p{2.5 cm}|p{2.5 cm}|}
\hline
\textbf{Sr. No} & \textbf{Research Question}& \textbf{Objective} & \textbf{Categorization }                          & \textbf{Relevant section} \tabularnewline
\hline
RQ1& What is the current state of Serverless Edge computing?  & It helps to understand serverless edge computing. The research question studies various areas of serverless edge computing.  & Analysis of primary studies and outcomes & \ref{Analysis}                 \tabularnewline
\hline
RQ2    & Which tools are available to implement or simulate serverless edge computing techniques? & The main objective of this question is to explore the tools used in the simulation or real-time implementation of serverless edge computing & Tool & ~\ref{Current Trend}              
\tabularnewline
\hline
RQ3 & What are the primary sub-areas in serverless edge computing?& Serverless edge computing can be identified by this question& Taxonomy  & ~\ref{Taxonomy}                 \tabularnewline
\hline
RQ4& What is the current state of serverless edge computing frameworks?                       & There are various designs developed for serverless edge computing.  This question investigates the different serverless frameworks & Frameworks&~\ref{Current State} \tabularnewline
\hline
RQ5& What kind of performance parameters are used in serverless edge computing frameworks?    & Different researchers used different performance parameters to evaluate the performance of serverless edge computing frameworks, such as energy, cost, etc. This question studied these parameters & Metrics& ~\ref{Current State}                 \tabularnewline
\hline
RQ6    & Which communication mode is used in serverless edge computing framework? & This question identifies the hardware in this study.& Tool &~\ref{Current Trend} \tabularnewline
\hline

RQ7  & What is the main driver of a serverless edge computing framework? & The driver is related to the work belonging to industry, academia, research, and development. & Tool&~\ref{Current Trend} \tabularnewline
\hline
RQ8   & What are the target applications of serverless edge computing frameworks?& Serverless may be designed for specific applications. This question helps to identify those applications& Application   &~\ref{Current Trend}                 \tabularnewline
\hline
RQ9   & What are the insights on the development of serverless edge computing& Analysis of existing research in the area of serverless edge computing& Results and Discussion&~\ref{results and discussion}                  \tabularnewline
\hline
RQ10   & What is the future of serverless edge computing?& This question includes research trends and challenges of SEC.     & Result and Discussion                    &~\ref{Current Trend}
\tabularnewline
\hline
\end{tabular}
\end{table*}
\end{center}
 \subsection{Information Sources}\label{Information Sources}
While performing a systematic literature review, it is important to consider broad searches of electronic resources~\cite{brereton2007lessons-28},\cite{stapic2012performing-31},\cite{kitchenham2012systematic-32}.  We have selected the following databases to enhance the chance of findings research articles relevant to our topic.
\begin{itemize}
    \item IEEE Explorer (\href{https://ieeexplore.ieee.org/Xplore/home.jsp}{ieeexplore.ieee.org} )
    \item Science Direct (\href{https://www.sciencedirect.com/}{www.science-direct.com})
    \item Springer Link (\href{https://link.springer.com/}{springerlink.com})
    \item ACM Digital Library (\href{https://dl.acm.org/}{dl.acm.org})
    \item Wiley Inter-science (\href{https://onlinelibrary.wiley.com/}{onlinelibrary.wiley.com})
    \item Google Scholar (\href{https://scholar.google.com/}{www.googlescholar.com})
\end{itemize}
\textbf{Additional Source:} We also consider additional sources for our systematic literature review, such as technical reports, edited books, textbooks, etc.
\subsection{Search Strategy}\label{Search Strategy}
Table~\ref{tab: Search String} indicates the search strategy for our resources. In this review, well-known online scientific resources are used to collect the research papers. IEEE Explorer, Science Direct, Springer, ACM Digital Library, Wiley inter-science, and  Google Scholar are the most commonly used digital libraries for retrieving articles~\cite{jatoth2015computational-33},\cite{vakili2017comprehensive-34}. Finding relevant research articles in the literature relies heavily on the "Search string creation" and "Search keywords selection"~\cite{riaz2010experiences-35},\cite{batool2022software-36}. The collection of search strings is used to describe serverless edge computing, frameworks, applications, and architecture. To get the final result, we combined the keywords using the Boolean operators AND and OR~\cite{asghari2019internet-37}.

("Serverless edge computing" OR "Serverless Edge Computing (SEC)"),
("Edge Computing" OR "Edge Computing (EC)") AND ("Framework" OR "Platform" OR "Application" Or "Architecture")
\begin{table*}[]
\caption{Search Strings for E-Source}
\label{tab: Search String}
\begin{tabular}{|p{2.5 cm}|p{2.5 cm}|p{2.5 cm}|p{3.5 cm}|p{2.5 cm}|}
\hline
\textbf{SR.No} & \textbf{E \_Source} & \textbf{Source Type}                     & \textbf{Search Title} &\textbf{Number of Studies}                                   \\ \hline
1              & IEEE Xplorer        & Journal, Conference,Magazine, Transaction & Serverless edge computing,Serverless edge, Edge computing& 34 \\ \hline
2              & Science Direct      & Journal, Conference                      & Serverless edge computing,Serverless edge, Edge computing& 3\\ \hline
3              & Springer Link      & Journal, Conference                    & Serverless edge computing,Serverless edge, Edge computing& 1\\ \hline
4              & ACM Digital Library & All Sources                              & Serverless edge computing,Serverless edge, Edge computing & 4 \\ \hline
5              & Wiley Inter Science & Journal, Conference                      & Serverless edge computing,Serverless edge, Edge computing& 1 \\ \hline
6              & Google Scholar      & All Sources                              & Serverless edge computing,Serverless edge, Edge computing& 4\\ \hline
\textbf{Total Studies}&&& &\textbf{47}\\ \hline
\end{tabular}
\end{table*}
\subsection{ Inclusion and Exclusion}\label{Inclusion and exclusion}
Serverless edge computing is quite a new area in research, and very few articles have been found in this field. Figure~\ref{fig: keyword image} illustrates the selection process of our research studies. Firstly we found 192  papers in this area; after applying inclusion and exclusion, only 47 papers are left. These research articles are closely related to serverless edge computing. Most papers we found after 2017. The  explanation of each paper based on taxonomy is shown in the section

 Inclusion and Exclusion criteria are based on assumptions which are shown in Table~\ref{tab:InandEX}
 \begin{table}[]
\caption{Inclusion And Exclusion Criteria}
\label{tab:InandEX}
\begin{tabular}{|l|p{4.5 cm}|}
\hline
\textbf{Inclusion Criteria} &                                                                             \\ \hline
IC1                         & Selected studies should be Journal or conference papers                       \\ \hline
IC2                         & Selected studies should be focused on serverless edge computing or serverless \\ \hline
IC2                         & Selected studies must be in English                                           \\ \hline
\textbf{Exclusion Criteria} &                                                                               \\ \hline
EC1                         & Studies which are not related to serverless edge computing.                   \\ \hline
EC2                         & Studies which are not in English.                                             \\ \hline
\end{tabular}
\end{table}
\subsection{Quality assessment}\label{Quality assessment}
serverless edge computing is a new research area, and we do not find so many articles in this field. We selected the most related articles. after performing inclusion and exclusion criteria. We conducted a quality evaluation using "The Centre for Reviews and Dissemination (CRD) guidelines"~\cite{kitchenham2004procedures-11}  to evaluate essential criteria, including external and internal validity and examined them for bias.
\subsection{Data Extraction}\label{Dataextraction}
-- showed the strategy for all 48 primary studies selected in this Systematic Literature review. To respond to the queries that were intended, we developed this data extraction form at the time when the data-gathering operation was started. Inclusion and exclusion criteria are clearly defined Table~\ref{tab:InandEX} which helped us find the articles related to serverless edge computing. Journal and conferences listing is also shown in--.  Data extraction rules are as follows:
\begin{itemize}
    \item All 47 primary studies were reviewed by one author.
    \item Another author used random samples to verify the accuracy of data collection.
    \item Conflicts that arose throughout the cross-checking process were discussed and settled in several different sessions.
\end{itemize}

\section{Current State of Serverless Edge Computing}\label{Current State}
\subsection{ Analysis of Selected Studies}\label{Analysis}
This section is based on primary studies, which include four categories such as development, metrics, frameworks, and platforms. Figure~\ref{fig: taxonomy} illustrates the taxonomy of serverless edge computing.
\begin{figure}[h]
    \centering
    \includegraphics[width= 8cm]{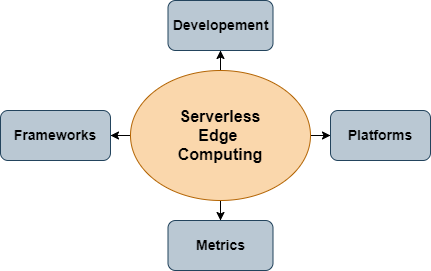}
    \caption{Taxonomy of serverless Edge computing}
    \label{fig: taxonomy}
\end{figure}

\cite{cicconetti2020decentralized-5} proposed a decentralized framework for serverless edge computing, which is gaining interest for IoT applications. The authors propose a framework for the effective dispatching of tasks in the network while reducing the response time. The choice is based on weights local to every e-router; nevertheless, the SDN controller is alerted of any congestion events. Moreover, the solution was lightweight and could be deployed on devices with limited resources, such as IoT gateways. Overall, for their study, they designed three algorithms. The results indicate that their design can easily manage rapidly changing load and network circumstances. However, a two-tier overlay is good for reducing computational requirements while maintaining or improving performance over a flat overlay. 
\cite{cicconetti2018architectural-41} leverages the software-defined network (SDN) for serverless edge computing. The authors study the challenge of dynamic allocation of stateless computations such as lambda functions. Their study developed an architecture for UTs to outsource the execution of stateless distant processes to an edge network. The entry points are e-routers that route lambda functions to the e-computers considered most suitable at the time. The selection is made based on local weights that are maintained in an up-to-date state by a separate process, which also interacts with the network controller of the SDN-enabled network that lies under the surface. Furthermore, they developed a tool for evaluating the performance of edge computing via the use of Mininet emulation. This tool allows edge devices to connect using an RPC interface.

\cite{mohanty2018evaluation-16} compared features of the open-source serverless computing frameworks. In their study, they use three open-source frameworks such as Fission, Kubeless, and OpenFaaS. They investigated that OpenFaaS is more flexible for supporting multiple container orchestration. To evaluate the performance, they deployed all three open-source frameworks on the Kubeless cluster. They categorize the performance in response time and success ratio. However, Kubeless's performance was better compared with other frameworks.
Similarly \cite{palade2019evaluation-42} investigate four open-source serverless platforms such as Kubeless, Apache OpenWhisk, OpenFaaS, and Knative. Each framework is deployed on a Kubernetes cluster with a bare-metal, single-master deployment. The authors showed some typical scenarios where these frameworks can be used. They outlined a list of criteria that a serverless computing framework may need to supply to allow the possible features of serverless computing at the edge of the network. They investigated that Kubeless outperforms compared to OpenWhisk. In another study, author~\cite{das2018edgebench-57} presents an edge benchmark for serverless edge computing. They studied two edge platforms, Greengrass and Azure Edge. According to the findings of their research, the performance of Greengrass and Azure Edge is similar, with the caveat that Azure Edge has greater end-to-end latency than Greengrass does because of its batch-based processing methodology.~\cite{rajput2022edgefaasbench-74} developed EdgeFaasBench,  a benchmark for serverless edge computing on edge devices. It is developed on top of Apache OpenFaaS with Docker Swarm and can run several workloads on edge nodes with different hardware. EdgeFaaSBench can gather a wide variety of metrics, both traditional and serverless-specific, that indicate the performance of edge devices. EdgeFaaSBench can collect cold and warm start times of serverless applications and performance degradation with concurrent function execution.

\cite{glikson2017deviceless-10} proposed a device-less edge computing paradigm. The authors conducted a series of experiments to realize device-less edge computing.~\cite{ko2022function-47} proposed a function-aware framework for serverless edge computing. The original purpose of the framework is to determine which instances should be maintained as a warm state. By formulating and solving a constrained Markov decision process problem on the joint optimal policy on which container instances are to be maintained under a warm status and how many CPU cores are assigned to them, this framework achieved high resource utilization while guaranteeing on-time task completion.~\cite{hall2019execution-68} provide a taxonomy for classifying serverless function access patterns, which enables the extraction of the fundamental needs of a serverless computing runtime. They present WebAssembly as an alternative method for running serverless applications. They conclude that a serverless platform built on WebAssembly may provide many of the same isolation and performance guarantees as container-based systems while decreasing typical application start times and the resources required to host them. 

\cite{yadav2021browser-79} developed and implemented an in-browser real-time facial recognition-based attendance tracking system employing tensorflow.js, face-api.js, and Google Cloud for dynamic storage in real-time. The system makes use of serverless edge computing in the process of updating attendance data using the Google Sheets API to save them on the fly, which helps to reduce the total latency experienced during the process. In another study, author~\cite{wang2021wearmask-45} proposed a solution for face mask detection, which is based on in-browser serverless edge computing. Wear-Mask can be deployed on any device, such as mobile phones, tablets, laptops, etc. The authors provide a comprehensive edge-computing framework by merging (1) deep learning models (YOLO), (2) a high-performance neural network inference computing framework (NCNN), and (3) a stack-based virtual machine the contribution that the proposed technique makes (WebAssembly). Inadequate support for deep learning from the JavaScript community was addressed by the deployment method, which aggregated NCNN and WASM. WebAssembly is a potentially useful alternative for doing serverless configuration at the edge.~\cite{mendki2020evaluating-66}. It allows for a quicker startup and more secure, sand-boxed execution, which are both benefits.

\cite{baresi2019towards-9} presents a framework for managing large-scale edge technologies. PAPS uses the serverless computing concept and uses a hierarchy of three control loops to manage its operations. It divides the edge topology into autonomous, delay-aware communities that are controlled by a central authority. The results showed that the approach performs very well under extreme workloads.~\cite{lin2021privacy-52} presents a serverless edge learning framework as a means of improving the effectiveness of distributed training from the point of view of networking. The framework can accomplish dynamic orchestration to use networking system resources while taking into consideration the heterogeneity of the underlying infrastructure and the learning tasks that are being requested.
\cite{pfandzelter2020tinyfaas-51} proposed a tinyFaas framework that is specifically designed for edge environments. To demonstrate the viability of their proposed solution, they built a proof-of-concept prototype. They conducted a series of experiments that directly compared their prototype's performance to that of Lean OpenWhisk, the only edge-focused open-source FaaS platform. Overall, tinyFaas performs better as compared to other open-source platforms .~\cite{cicconetti2022faas-65} investigates the possibility of supporting stateful applications on the serverless systems that are distributed on the edge nodes. In their study, they concentrated on the issue of moving the state along an invocation of functions in chain and DAG workflows. As a result, we have uncovered three potential alternative schemes, each of which has distinctive qualities. They designed a prototype implementation to demonstrate the practicability of their techniques and to test performance while accounting for actual protocol overheads. According to the findings, passing along the current state as one moves from one function call to the next may drastically reduce the amount of communication overhead.  

\cite{lin2021privacy-52} presents a serverless edge framework to improve the network efficiency of distributed training. The framework can accomplish dynamic orchestration to use networking system resources while taking into consideration the heterogeneity of the underlying infrastructure and the learning tasks that are being requested. 
The open issues and potential future research paths of serverless edge learning are presented in this article to inspire the development of effective distributed training systems.~\cite{nastic2017serverless-6} author introduces a serverless edge data analytic platform and discusses their main design and challenges. The serverless stream architecture that was presented makes the underlying heterogeneous edge infrastructure visible, which allows for the quicker and more intuitive construction of a variety of real-time analytics tasks without the need to worry about nonfunctional needs.~\cite{deng2021dependent-55} studied the dependent function embedded problem at the serverless edge. They begin by pointing out that proactive stream mapping and data splitting might potentially have a significant effect on the makespan of DAGs with many use cases. The DPE method, which we designed based on these facts,  has been theoretically confirmed to reach global optimality for any arbitrary DAG. DPE can determine the ideal stream mapping for any function pair that has dependent relations.

\cite{aslanpour2022energy-54} proposed energy-aware for serverless edge computing. In this study, the researcher examines the issues associated with enhancing energy awareness at serverless edge computing using resource scheduling. They proposed resource-scheduling algorithms to place the function on the edge node. The proposed algorithm's objective is to increase the operational availability of edge nodes while reducing the variation thereof without compromising the throughput and QoS.~\cite{tang2022distributed-56} investigated the issues of distributed job scheduling in serverless edge computing networks for the Internet of Things. They formulated the optimization problem as a POSG because of the heterogeneous and resource-constrained nature of serverless edge computing nodes. In this model, multiple serverless edge computing nodes are non-cooperative in scheduling computation tasks from IoT clients and allocating computing resources to optimize their utility. The authors developed the multi-task scheduling algorithm to solve the partial observability issue. 

\cite{lakhan2022restricted-58} leverages the mobility-aware security dynamic service composition (MSDSC) algorithmic framework for workflow healthcare based on serverless, serverless, and constrained Boltzmann machine techniques. The effort resulted in the creation of modern approaches such as a security-aware complete homomorphism approach, service composition method, job sequencing method, and scheduling method. According to the results of the performance assessment, the proposed model and schemes are better than all current baseline methods in terms of both time and cost of execution. The main objective is to reduce the cost.

\cite{hussain2019serverless-59} proposed a serverless edge computing system to enable green oil extraction. To deal with the unknowns that come with operating in distant offshore smart oil fields, the system employs a robust resource allocation methodology. The authors made use of edge federation to utilize idle neighbor edge devices and prior data to develop a probabilistic model that could estimate when an incoming service request (task) would be completed.

\cite{carpio2021engineering-61} developed a serverless edge computing platform for serverless applications. The main objective is to provide storage, processing, and communication using only open-source software. Authors benchmark their system according to scalability, throughput, and response time. The results indicate that raspberry pi performs better. To concurrently optimize time, energy, and cost,~\cite{xie2022workflow-62} designed the workflow application as a directed acyclic graph (DAG) to reflect the association of the deconstructed serverless processes. They built a D3QN-based algorithm to solve the problem, which captured the volatility of the serverless edge computing system. The researcher performs various experiments to verify the effectiveness of the proposed solutions.
\begin{table*}[]
\caption{Comparison of SLR with existing Studies}
\label{tab: Comparison with existing studies}
\begin{tabular}{|p{2.5 cm}|p{4.5 cm}|l|l|l|l|l|l|l|l|l|l|l|}

\hline
\textbf{Research Paper}                                 &                            & \cite{aslanpour2021serverless-8}                     &\cite{ioini2020platforms-22}                    & \cite{xie2021serverless-15}                  & \cite{li2022serverless-38}         &\cite{kjorveziroski2021iot-48}                  & \cite{gadepalli2019challenges-49}                  & \cite{cassel2022serverless-20}                    & \cite{taibi2020patterns-24}                    &\cite{eismann2020review-25}                    & \cite{yussupov2019systematic-26}                    & \cite{scheuner2020function-27}                      \tabularnewline
\hline
\textbf{Taxonomy }                                      & Development                & \Checkmark & \Checkmark & \Checkmark & \Checkmark &                       &                       & \Checkmark & \Checkmark &                       & \Checkmark &                         \tabularnewline
& Latency \& reliability     &                       &                       &                       &                       &                       &                       &                       &                       & \checkmark &                       & \checkmark \tabularnewline

& Resource Optimization      &                       &                       &                       &                       & \Checkmark & \Checkmark &                       &                       & \Checkmark & \Checkmark &                         \\
& Security \& privacy        &                       &                       &                       &                       &                       &                       &                       & \Checkmark&                       & \Checkmark & \Checkmark \\
& Qos&\Checkmark&                       &                       &                       & \Checkmark &                       &                       & \Checkmark &                       &                       &                         \\
& Provisioning               &                       &                       &                       &                       &                       & \Checkmark & \Checkmark &                       &                       & \Checkmark &                         \\
& Energy \& Cost Consumption &                       &                       &                       &                       &                       &                       &                       & \Checkmark &                       &                       &                         \\
 & H/w and S/w                &                       &                       &                       &                       & \Checkmark &                       &                       & \Checkmark &                       &                       &                         \\
 & Application Support        &                       & \Checkmark& \Checkmark &                       &                       & \Checkmark & \Checkmark &                       &                       & \Checkmark&                         \\
 & Frameworks                 &                       & \Checkmark &                       &                       & \Checkmark & \Checkmark &                       &                       &                       &                       & \Checkmark   \tabularnewline
 \hline
\textbf{Comparison  }                                   & Optimal Parameters         &                       &                       & \Checkmark &                       & \Checkmark&                       &                       & \Checkmark &                       &                       &                         \\
& Open Issues                & \Checkmark & \Checkmark&                       &                       &                       & \Checkmark & \Checkmark &                       &                       &                       &                         \\
 & Evaluation Factors         & \Checkmark & \Checkmark & \Checkmark &                       &                       & \Checkmark& \Checkmark & \Checkmark &                       & \Checkmark & \Checkmark  \\
& SLR Based                  &                       &                       &                       &                       &                       &                       &                       & \Checkmark &                       &                       &                         \tabularnewline
\hline
\textbf{Future Direction for Serverless Edge Computing} &                            & \Checkmark & \Checkmark&                       & \Checkmark&                       & \Checkmark & \Checkmark& \Checkmark &                       & \Checkmark &   
\tabularnewline
\hline
\end{tabular}
\end{table*}
\cite{rausch2021optimized-63} proposed a container scheduling system named Skippy to enable existing container orchestration to support serverless edge functions. Skippy does this by imposing scheduling requirements that use supplementary information about node capabilities, the data flow of the application, and the network architecture. The author demonstrates that the trade-off between data movement and computational movement is an important factor to consider when determining the location of functions in data-intensive serverless edge computing.~\cite{yao2023performance-64} presents a deep reinforcement learning-based model (ES DR). The main objective of this model is to reduce the latency and the cost of IoT devices and increase their success rate. First, they formulate the optimization issues in serverless function offloading as MDP and then solve it by the DRL method. Simulation experiments are utilized to evaluate how well our strategy performs in comparison to four other methods that are already in use. The findings indicate that ES-DRL has the potential to successfully lower the average latency of functions, as well as the exploration cost of functions, while simultaneously increasing the success rate of functions.~\cite{jeon2021deep-73} present a deep reinforcement learning-based algorithm that effectively caches critical and popular packages with per-function response time QoS in hierarchical edge clouds. In this work, cooperative MARL is used to strike a compromise between hit rate and response time QoS by training each caching agent using MADDPG and a global reward that carefully accounts for both hit rate and the number of violations of response time QoS.

\cite{aske2018supporting-67} developed LAVEA, a low-latency video edge analytic system. This system interacts with local clients, edge, and distant cloud nodes, and it converts video inputs into semantic information at locations that are closer to the users at an earlier stage. To reduce the time, the researcher adopted the edge-front architecture and defined an optimization problem for offloading job selection.~\cite{coviello2022dataxe-71} present DataXe, which automatically scales the stateful and stateless data analytics units in an edge environment. For real-time video analytics, the performance of DataXe is 3 times faster than custom analytics applications.~\cite{lyu2022towards-69} investigates the feasibility of integrating support for DAG functions into Sledge, a 
 WASM-based serverless runtime. In their study, lightweight sandboxes are used  by Sledge, which provides the perfect environment for the development of function chains. Sledge aims to reduce the cost between the functions.

 \cite{sethunath2021joint-70} leverage an architecture that incorporates cross-monitoring, which allows for rapid responses when nearby Edges are in failing condition.  This design enables instantaneous responses by importing measurements from a nearby Edge straight to the Edge. In this way, the Zone's adjacent Edges may safely back up against one another.
 
\cite{cicconetti2021preliminary-72} propose a framework for distributing calls to FaaS-based apps at the network's periphery. QUIC is a protocol that has evolved from TCP and has various properties that make it suitable for usage in edge applications, particularly for mobile users. Results indicate that using QUIC+HTTP/3  in serverless is efficient for performance.~\cite{mistry2020demonstrating-85} explore the Unikernel as a way to construct the serverless platform at the edge with the main focus on security and performance. They introduce UniFaaS, a prototype edge-serverless platform that uses unikernels, which are small library single-address-space operating systems that only include the OS elements required to run a certain application to perform functions. The result indicates that a serverless platform that has very small footprints in terms of both memory and CPU and excels in performance.

\cite{aslanpour2022energy-54} examine the issues associated with enhancing energy awareness at Serverless edge computing using resource scheduling. They presented resource scheduling algorithms to arrange functions on edge nodes powered by a battery and renewable energy sources such as solar. The main objective of the algorithm is to increase the operational availability of edge nodes while simultaneously decreasing the fluctuation in that availability, all without affecting throughput or quality of service. The findings indicate that their proposed energy-aware scheduler has the potential to increase operational availability, throughput, and serviceability in comparison to the benchmark algorithms and get near to optimum while still preserving the quality of service.~\cite{smith2022fado-76} introduced FaDO: FaaS Functions and Data Orchestrator that accomplished data-aware function placement and automated granular data replication across heterogeneous serverless compute clusters. Utilizing buckets as both the units of replication and the measure for choosing function invocation locations, FaDO can use the storage bucket paradigm offered by cloud computing. Results indicate that FaDO's performance was very good. ~\cite {rausch2019towards-77} proposed a serverless platform that elevates concepts from the AI workflow to first-class citizens and provides a more approachable way to develop and operate edge AI functions. Their solution to device-less function scheduling considers device capabilities (such as newly introduced AI accelerators) and data locality. As a result, they were able to conceal from the developer difficult operational data and model management duties.

\cite{chaudhry2020improved-4} proposed a technique that uses serverless edge to combine MEC and NFV at the system level and developed VNFs on request. This is accomplished by merging MEC functional blocks with an NFV orchestrator that is run on a Kubernetes cluster. They studied if the resource usage of such an MEC system could be increased further by exploiting networked FPGA-enabled MEC nodes. This was done via an expansion in the MEC layer that used existing programmable hardware such as NetFPGA-10G. For experiments, they used two different scenarios, and in the end, the results were encouraging.~\cite{gill2021quantum-80}  discussed various paradigms such as serverless computing, serverless computing, quantum computing, and blockchain. The author indicates that with blockchain and quantum computing, serverless edge services can be reliable, and it also increases security and computational performance.

\cite{li2022kneescale-81} proposed KneeScale, an adaptive auto-scaler that effectively scales serverless services in edge systems under resource budget limitations. They designed a unique method for dynamically adjusting the number of function instances to swiftly reach a Knee point where the relative cost of increasing resource allocation is no longer worth the accompanying performance advantage. KneeScale was deployed on a Kubernetes-based OpenFaaS infrastructure, and its efficacy was assessed using FunctionBench, a standard FaaS benchmark. KneeScale outperforms previous auto-scaling algorithms under resource budget limitations for both fixed and dynamic workloads, according to experimental data.~\cite{paraskevoulakou2021leveraging-82} presents a way to achieve the supply of machine learning functions as a service, also known as ML-FaaS, by using the Apache OpenWhisk event-driven, distributed serverless platform. ML-FaaS stands for machine learning functions as a service. This method deals with composite services in addition to single ones. These composite services are workflows of machine learning jobs that include procedures such as aggregation, cleaning, feature extraction, and analytics. These workflows depict the whole data journey. The proposed approach was evaluated using a typical use case representing an ML pipeline.
\begin{center}
\begin{table*}
\caption{Comparison of existing evaluation factors with serverless edge computing framewrok} \label{tab:long}
\begin{tabular}{|p{1cm}|p{1.5 cm}|p{1 cm}|p{2 cm}|p{1 cm}|p{1 cm} |p{1 cm}|p{2 cm}|}
\hline
\multicolumn{1}{|c|}{\textbf{Studies}} & \multicolumn{1}{c|}{\textbf{Resource Management}} & \multicolumn{1}{c|}{\textbf{Latency}} & \multicolumn{1}{c|}{\textbf{Privacy \& Security}} & \multicolumn{1}{c|}{\textbf{Reliability}} & \multicolumn{1}{c|}{\textbf{QoS}} & \multicolumn{1}{c|}{\textbf{Cost \& Energy Consumption}} & \multicolumn{1}{c|}{\textbf{Frameworks}} \\ \hline

\cite{rausch2021optimized-63}&-& \checkmark &-&-&\checkmark& \checkmark& - \\ \hline
\cite{yao2023performance-64}& -& \checkmark& -& -& -& \checkmark& ES-DRL \\ \hline
\cite{cicconetti2022faas-65}&-  & \checkmark&-& \checkmark&-& -&- \\ \hline
\cite{das2018edgebench-57} &\checkmark&\checkmark& -& -& -& \checkmark& Amazon AWS Greengrass, Microsoft Azure \\ \hline
\cite{pfandzelter2020tinyfaas-51}&\checkmark&\checkmark&-&\checkmark & -& -& TinyFaaS \\ \hline
\cite{wang2021wearmask-45}& -& -& \checkmark& \checkmark& -& -&-\\ \hline
\cite{carpio2021engineering-61} &-&-&- &-&-&-& Decentralized Network Computing Platform \\ \hline
\cite{lin2021privacy-52} &-&-&\checkmark&\checkmark&-&-&Edge Learning Framework \\ \hline
\cite{glikson2017deviceless-10}  &\checkmark&-& \checkmark& -& -& -& Deviceless Edge Computing\\ \hline
\cite{aske2018supporting-67} &- &\checkmark &-& \checkmark&\checkmark& -&MPSC \\ \hline
\cite{hall2019execution-68}&-& \checkmark&- &\checkmark&\checkmark&-&WebAssembly based serverless Platform\\ \hline
\cite{lyu2022towards-69}&\checkmark&-&\checkmark&- &-&- &Sledge Wasm Based serverless Platform \\ \hline
\cite{cicconetti2020decentralized-40}& \checkmark&\checkmark&- &- &-& -&Open Whisk \\ \hline
\cite{sethunath2021joint-70}&- &\checkmark&- &-&\checkmark &- & - \\ \hline
\cite{cicconetti2021preliminary-72}&-& -& -&-&\checkmark &-& QUIC \\ \hline
\cite{nastic2017serverless-6}& \checkmark& -& -& \checkmark& -& -& Real Time Analytic Platform \\ \hline
\cite{cicconetti2018architectural-41} &-& \checkmark & -& \checkmark&- & - &SDN Enabled Network Architecture \\ \hline
\cite{palade2019evaluation-42}&-&\checkmark& -& -&\checkmark& - &Kubeless, Apache OpenWhisk, Apache FaaS, Knative\\ \hline
\cite{mistry2020demonstrating-85} &-&\checkmark &- &\checkmark&\checkmark&-&UniFaaS \\ \hline
\cite{deng2021dependent-55}  & - &  \checkmark  & -  & -& - & -&- \\ \hline
\cite{tang2022distributed-56}  &-&-&-& \checkmark &\checkmark&- & - \\ \hline
\cite{rajput2022edgefaasbench-74} &-&- & -& \checkmark& \checkmark & -& EdgeFaasBench, Apache OpenFaas \\ \hline
\cite{smith2022fado-76}& - & - & - & \checkmark& -& \checkmark& FaDo \\ \hline
\cite{yadav2021browser-79}  & -&-&\checkmark& \checkmark & \checkmark&-&  - \\ \hline
\cite{ko2022function-47}& \checkmark & - & -&  & \checkmark& \checkmark & FARM\\ \hline
\cite{rausch2019towards-77}& -& -&-&-&\checkmark & - &- \\ \hline
\cite{baresi2019towards-60} &- &\checkmark&- &- &\checkmark& -& FaaS \\ \hline
\cite{huang2022mobility-83}& - &-&-  &-& \checkmark&-&-\\ \hline
\cite{mohanty2018evaluation-16}& -&- & -& \checkmark& -&-& Fission, OpenFaaS, Kubeless, OpenWhisk \\ \hline
\cite{aslanpour2022energy-54} & \checkmark &- &-& \checkmark& \checkmark &\checkmark&-\\ \hline
\cite{li2022joint-88}&-  & -& -& -&\checkmark & -&  - \\ \hline
\cite{pelle2020operating-91}  &-& \checkmark&-& -& \checkmark& \checkmark & Amazon AWS, Lambda, Greengrass \\ \hline
\cite{lakhan2022restricted-58} & \checkmark & -& \checkmark &- & \checkmark& \checkmark & MSDSC \\ \hline
\cite{kjorveziroski2022kubernetes-92}&- &\checkmark & -& \checkmark& -& -& Open FaaS Serverless\\ \hline
\cite{bac2022serverless-93}& - & - & - & \checkmark &-& - & - \\ \hline
\cite{hussain2019serverless-59} & - & -& -&  - &\checkmark & - &  - \\ \hline
\cite{zhang2020stoic-95} & \checkmark &-&-&-&\checkmark&-& STOIC \\ \hline
\cite{tutuncuouglu2022online-90} & - & - & - & - & \checkmark &  - &- \\ \hline
\cite{xie2022workflow-62}& \checkmark&- & - &-& - & \checkmark &  -  \\ \hline

\end{tabular}
\end{table*}
\end{center}

\cite{huang2022mobility-83} presented a device-less edge computing system focused on end-user mobility. They designed a smooth virtual function movement strategy to serve customers with continuous services when they are mobile. The researchers created a migration interface to facilitate function transfer while needing little input from function developers and attaining user transparency. They developed a streaming game as a typical virtual function for packet-intensive and delay-sensitive circumstances and deployed it on real-world edge devices. They conducted several experiments to compare their proposed algorithm with other algorithms. The findings reveal that the system performs effectively under actual virtual function and user mobility, offering smooth virtual function migration and significantly improving user experiences when compared to existing methods.~\cite{baresi2019towards-9} deployed NEPTUNE, a serverless-based solution for managing latency-sensitive applications. It includes intelligent placement and routing to reduce network overhead, dynamic resource allocation to respond to changes in workload swiftly, and transparent control of CPUs and GPUs. All of these features help to reduce the amount of time spent on managing the network. They illustrated the practicability of the method and got intriguing results in comparison to other state-of-the-art solutions with the assistance of a prototype that was created on top of K3S, which is a popular container orchestrator for the edge.

\cite{li2022joint-88} examined how to speed up the performance of serverless applications in edge clouds by selecting the best communication styles and assigning the relevant functions to those functions. They presented the joint optimization problem for achieving the shortest possible application completion time in the form of a nonlinear integer programming problem, and we demonstrated that it is an NP-hard issue. In addition to this, they presented a heuristic method known as PASS, the approximation ratio of which is theoretically investigated. They thoroughly validate the feasibility of the PASS method using trace-driven simulations, and the results show that it truly greatly outperforms other current state-of-the-art function assignment algorithms.~\cite{tutuncuouglu2022online-90} investigated the interplay between latency-constrained services and resource management for serverless edge computing, and we provided a modelling abstraction and a problem formulation to do so. Based on a queuing network model of task graph execution, the provided abstraction allows for the study of the interaction between selfish WDs that reserve edge resources and a serverless operator that distributes resources among WDs, presented as a non-cooperative game. Analytical findings reveal that rate reservation is critical for latency-sensitive services, while a simple abstraction for rate reservation enables conceptually simple algorithms, such as the proposed OARC, to converge to a good-performing equilibrium.

\cite{pelle2020operating-91} followed the cloud-native programming
 and serverless operating techniques for latency-sensitive IoT
 applications. A unique architecture was presented on top of the cloud platform that provided serverless solutions. The overall method was applied to Amazon's AWS, using its FaaS services, Lambda and Greengrass. They indicate the cost and latency were significantly affected by the constituent functions.~\cite{pan2022retention-92} investigate the retention-aware container caching issue in serverless edge computing to enhance system efficiency by combining container caching with request distribution. They demonstrate via analysis that this issue is difficult to solve even when simplified. We show that in certain situations, the issue may be translated to the traditional ski-rental problem, and we present an online competitive method for these cases.~\cite{kjorveziroski2022kubernetes-92} compared the performance of three distinct Kubernetes distributions: full-fledged Kubernetes, K3s, and MicroK8s. Findings indicate that lighter Kubernetes versions perform better on bare-metal devices when complex integration with third-party cloud platforms is unnecessary~\cite{bac2022serverless-93} introduced the serverless edge computing strategy for implementing machine learning applications in this study. Overall, the strategy demonstrated serverless computing's applicability for machine learning tasks done in distributed edge environments. They thoroughly assess the performance of a well-known case study image classification using the MNIST dataset throughout the training and serving processes. Their findings showed that the serverless solution could provide optimum response and operating time, decrease machine learning application end-to-end latency, and allow distributed machine learning.

 \cite{hussain2019serverless-59} presented a federation of serverless edge computing systems capable of green oil extraction. With a probabilistic model, they used the edge federation to use underused neighbour edge devices and historical data to anticipate the completion time of an incoming service request (task). The unknowns, such as those involving communication and calculation, were factored in throughout the process of estimating the likelihood of achievement within a certain time frame for the completion of the work. The findings of their experiments indicate that our suggested model has the potential to increase the percentage of times that urgent services are completed in comparison to other standard models.~\cite{zhang2020stoic-95} presented a framework named STOIC, executing machine learning applications in a hybrid cloud setting. Edge Controller, Edge Cloud, and Public Cloud were the three main components of STOIC. They indicate that STOIC can effectively plan machine learning jobs across hybrid cloud deployments and gain higher performance than any single deployment option in isolation. This is made possible by STOIC's ability to schedule machine learning activities intelligently.
 \begin{center}
\begin{table*}
\caption{Comparison of serverless edge computing in terms of Drivers, Applications, Comm Mode, Types and Sources} \label{tab:Comlong}
\footnotesize
\begin{tabular}{|l|p{3.5cm}|p{2.9cm}|p{3.5cm}|p{1.9cm}|p{2.5cm}|}
\hline 
\multicolumn{1}{|c|}{\textbf{Studies}} & \multicolumn{1}{c|}{\textbf{Drivers}} & \multicolumn{1}{c|}{\textbf{Applications}} & \multicolumn{1}{c|}{\textbf{Comm Mode}} & \multicolumn{1}{c|}{\textbf{Type}} & \multicolumn{1}{c|}{\textbf{Source}} \\ \hline 

\cite{rausch2021optimized-63} & Research \& Development & Other & Serverless Edge Computing & Simulator & Not Mentioned \\ \hline
\cite{yao2023performance-64} & Research \& Development & IoT Applications & IoT serverless Edge Computing & Event driven & Not Mentioned \\ \hline
\cite{cicconetti2022faas-65} & Research \& Development & DAG modeled Applications & Edge network & Network driven & Open Source \\ \hline
\cite{pfandzelter2020tinyfaas-51} & Research \& Development & IoT Applications & Edge Computing & Other & Not Mentioned \\ \hline
\cite{wang2021wearmask-45} & Research \& Development & All Devices & Serverless Edge Computing & Other & Not Mentioned \\ \hline
\cite{carpio2021engineering-61} & Academia & Serverless Application & Edge Computing & Network driven & Open Source \\ \hline
\cite{lin2021privacy-52} & Research \& Development & Mobile, Edge and Cloud Devices & Serverless Edge Computing & Distributed Training & Not Mentioned \\ \hline
\cite{hall2019execution-68} & Research \& Development & IoT Applications & Edge network & Network driven & Open Source \\ \hline
\cite{lyu2022towards-69} & Industry & DAG Applications & Serverless & Other & Open Source \\ \hline
\cite{cicconetti2020decentralized-40} & Research \& Development & IoT Applications & Edge & Network driven & Not Mentioned \\ \hline
\cite{sethunath2021joint-70} & Research \& Development & Serverless Applications & Edge network & Other & Not Mentioned \\ \hline
\cite{cicconetti2021preliminary-72} & Academia & Serverless Applications & Edge Network & Other & Not Mentioned \\ \hline

\cite{coviello2022dataxe-71} & Research \& Development & Video Analytic Applications & Standalone & Other & Not Mentioned \\ \hline
\cite{jeon2021deep-73} & Research \& Development & Other & Serverless Edge Computing & Other & Not Mentioned \\ \hline
\cite{mistry2020demonstrating-85} & Research \& Development & Serverless Applications & Edge & Other & Open Source \\ \hline
\cite{deng2021dependent-55} & Research \& Development & Other & Edge & Other & Open Source \\ \hline
\cite{tang2022distributed-56} & Research \& Development & IoT Applications & Serverless Edge Computing & Simulator & Not Mentioned \\ \hline
\cite{rajput2022edgefaasbench-74} & Academia & Edge Devices & Serverless Computing & Other & Open Source \\ \hline
\cite{mendki2020evaluating-66} & Industry & Container-based Applications & Serverless Edge Computing & Other & Not Mentioned \\ \hline
\cite{smith2022fado-76} & Research \& Development & Other & Serverless Computing & Data ware functions & Not Mentioned \\ \hline
\cite{yadav2021browser-79} & Academia & Other & Serverless Edge Computing & Real time & Not Mentioned \\ \hline
\cite{ko2022function-47} & Research \& Development & Serverless Edge & Edge Cloud & Other & Not Mentioned \\ \hline

\cite{gill2021quantum-80} & Academia & IoT Applications & Edge Computing & Other & Not Mentioned \\ \hline
\cite{li2022kneescale-81} & Academia & IoT Applications & Serverless Computing & Other & Open Source \\ \hline
\cite{paraskevoulakou2021leveraging-82} & Academia & ML Applications & Serverless & Other & Not Mentioned \\ \hline

\cite{huang2022mobility-83} & Academia & Mobile Application & Deviceless Edge Computing & Other & Not Mentioned \\ \hline
\cite{aslanpour2022energy-54} & Research \& Development & IoT Applications & Serverless Edge & Event driven & Open Source \\ \hline
\cite{li2022joint-88} & Academia & Serverless Applications & Serverless & Other & Not Mentioned \\ \hline
\cite{pelle2020operating-91} & Research \& Development & IoT Applications & Serverless Edge Cloud & Other & Open Source \\ \hline
\cite{lakhan2022restricted-58} & Research \& Development & Healthcare Applications & Serverless Edge Computing & Simulator & Not Mentioned \\ \hline
\cite{kjorveziroski2022kubernetes-92} & Industry & Serverless Applications & Edge network & Network driven & Open Source \\ \hline
\cite{bac2022serverless-93} & Academia & ML Applications & Serverless Computing & Other & Open Source \\ \hline
\cite{zhang2020stoic-95} & Academia & ML Applications & Edge Cloud & Other & Not Mentioned \\ \hline
\cite{tutuncuouglu2022online-90} & Research \& Development & Other & Serverless Edge & Semi simulator & Not Mentioned \\ \hline
\cite{xie2022workflow-62} & Industry & IoT Applications & Serverless Edge & Other & Not Mentioned \\ \hline

\end{tabular}
\end{table*}
\end{center}

\section{Results and Discussion}\label{results and discussion}
 In this section, we discuss the proposed taxonomy of serverless edge computing frameworks and trend analysis.
 \subsubsection{Taxonomy of Serverless edge computing}\label{Taxonomy}
In this section, a detailed taxonomy of serverless edge computing is proposed based on the existing systematic literature review. The taxonomy of serverless edge computing is shown in Figure~\ref{fig: Taxonomy of SEC} which includes development, metrics, platform, driver communication mode, type, source, and target applications. For a thorough investigation of serverless edge computing, each taxonomy parameter is subdivided into other categories. Recent literature reviews only concentrate on serverless edge computing opportunities and open issues. The chosen research publications that address and explain the frameworks of serverless edge computing are compared using the proposed taxonomy. The presented taxonomy also classified research articles according to criteria such as applications, development processes, kinds, and performance measures, among other things. In addition, the chosen articles are evaluated in terms of comparing several factors, including benefits, primary context, flaws, and end outputs.

\begin{figure*}
    \centering
    \includegraphics{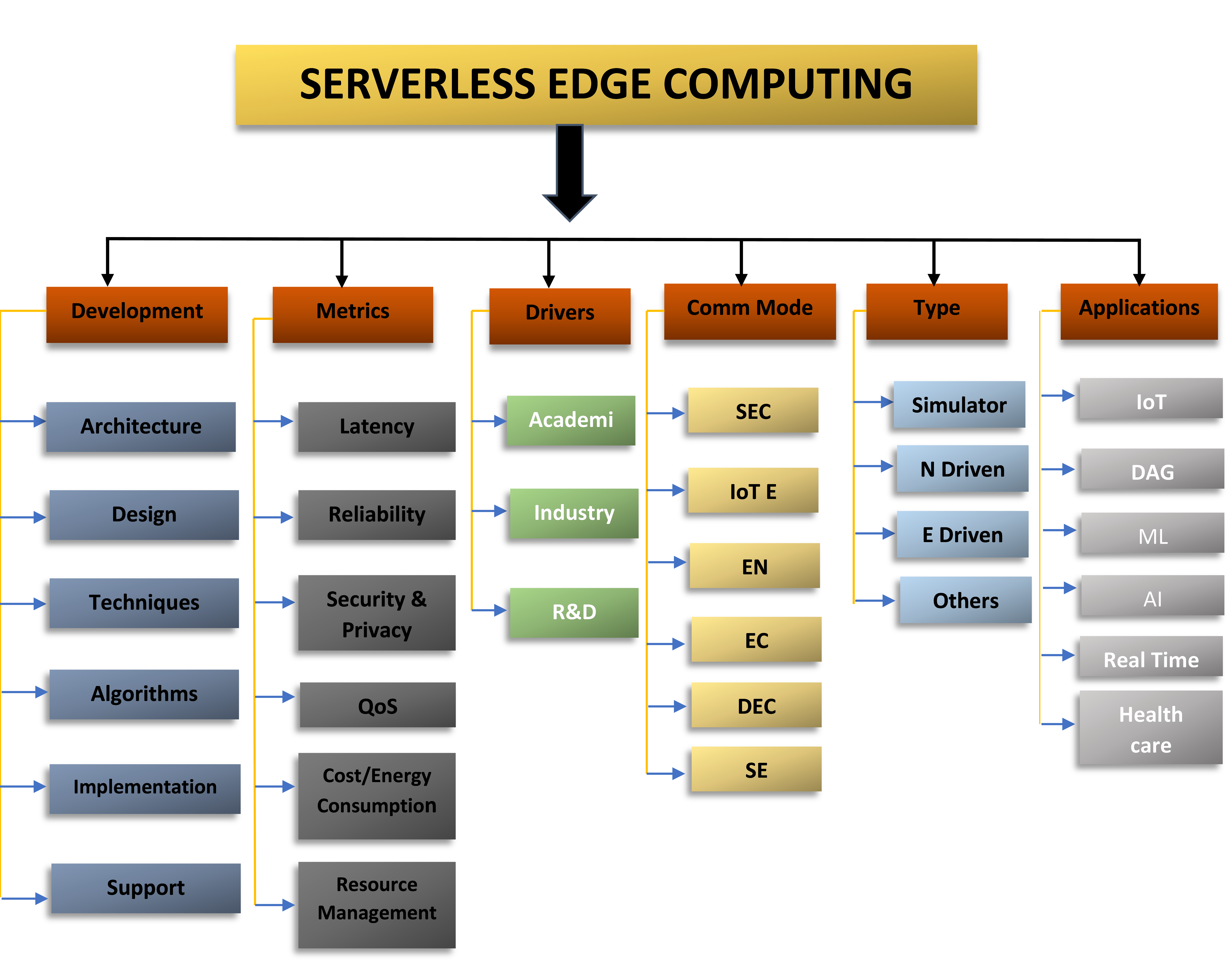}
    \caption{Taxonomy of Serverless Edge Computing}
    \label{fig: Taxonomy of SEC}
\end{figure*}
\subsubsection{Trend analysis and Observation}
Trends of serverless edge computing are shown in Figure~\ref{fig: distribution}, which is a year-by-year division depending on the publisher that is accessible from the overall selected research papers. Appendix~\ref{tab: Conference and Journal} illustrates the list of journals and conference papers on serverless edge computing. Trends indicate that the number of papers on serverless edge computing has increased recently. Figure~\ref{fig: Taxonomy of SEC} shows the important factors related to serverless edge computing in terms of resource management, latency, security and privacy, quality of services, and energy/cost consumption. Figure~\ref{fig: Papers}shows the year-by-year publication of research papers from the year 2017 to 2022. Figure~\ref{fig: Drivers} shows the distribution of research articles based on main drivers, communication mode, type application, and source.
\begin{figure*}
    \centering
    \includegraphics{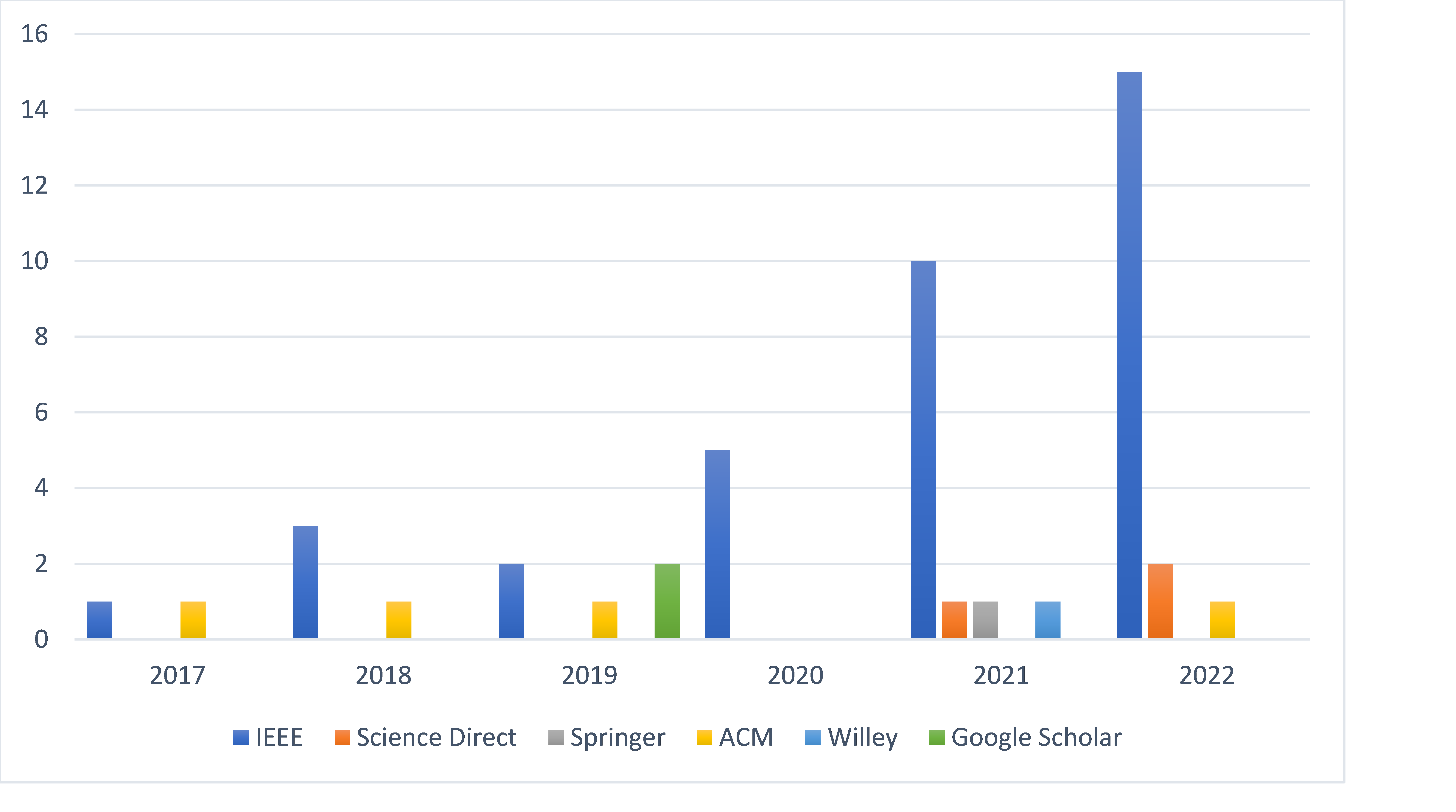}
    \caption{Distribution of research papers}
    \label{fig: distribution}
\end{figure*}

\begin{figure*}
    \centering
    \includegraphics{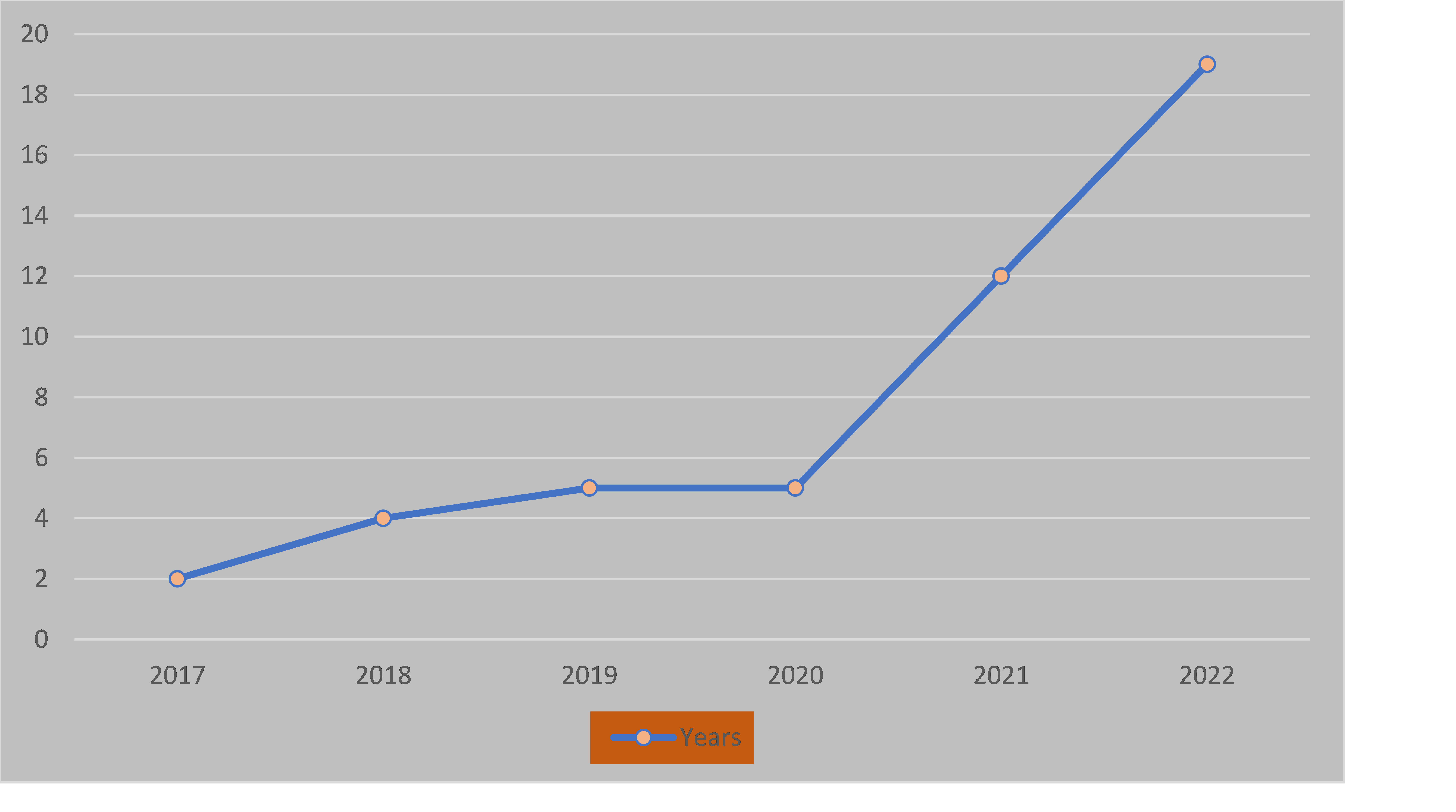}
    \caption{year-wise publications of serverless edge computing}
    \label{fig: Papers}
\end{figure*}

It has been observed that the trend for publishing research papers on serverless edge computing has been increasing year by year since 2017.  Considering the important factors related to serverless edge computing,  such as resource management, latency, reliability,  and applications of serverless edge computing, more research has been done.  We identify the quality of service parameters, such as response time, reliability, throughput, footprint, etc., that got the most attention. Security and privacy play a significant role in implementing the framework for serverless edge computing. However, energy and cost also play a vital role in developing serverless edge computing frameworks.
\begin{figure*}
    \centering
    \includegraphics{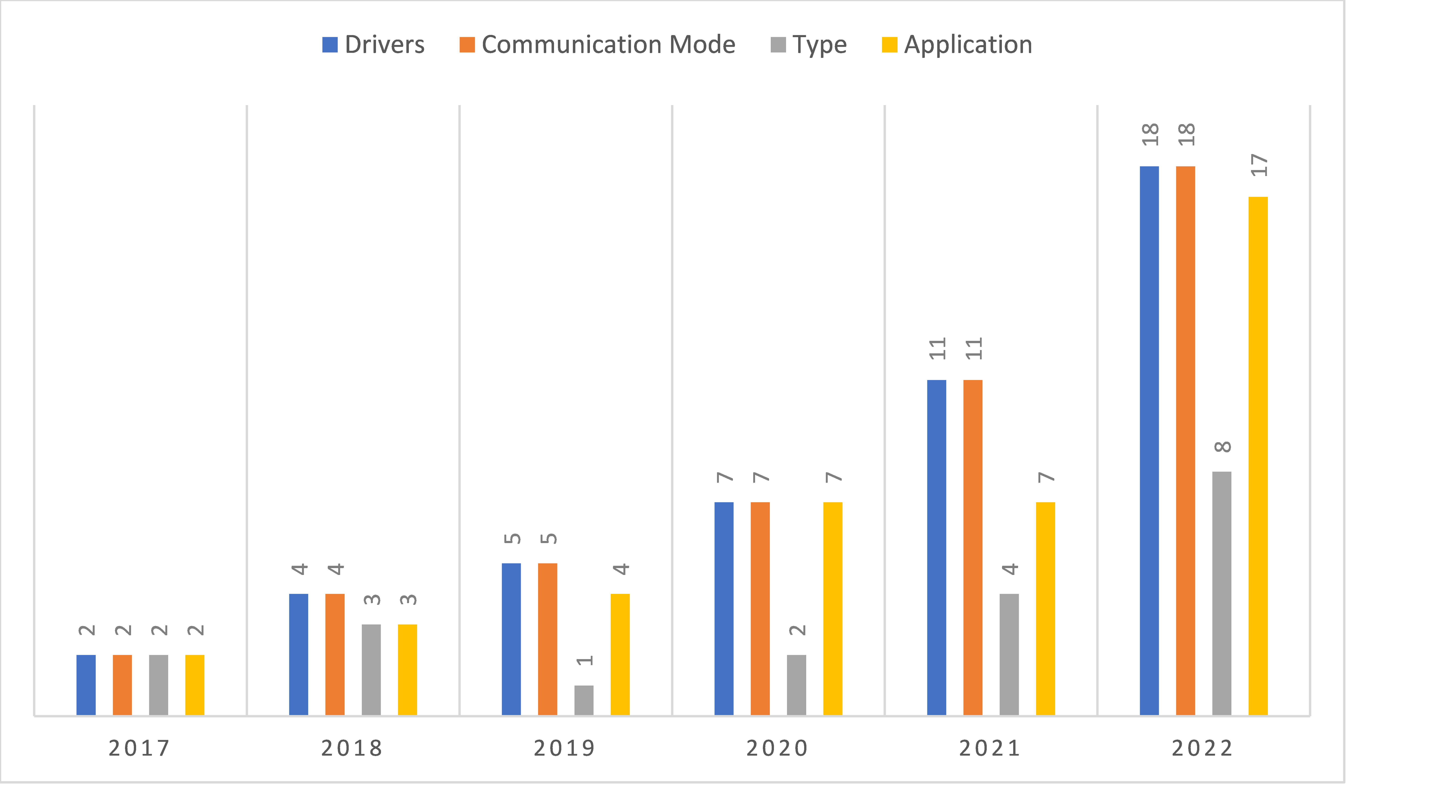}
    \caption{Distributions of papers based on Drivers, Communication Mode, Application and Type}
    \label{fig: Drivers}
\end{figure*}

\section{Current Trend and Future Directions}\label{Current Trend}
\subsection{Infrastructure and Architecture}
The technology of serverless edge computing is going to evolve in a manner that will make it necessary to get market adoption to support all smart devices. There are many challenges present in serverless edge computing~\cite{aslanpour2020performance-17},\cite{xie2021serverless-15}.

\subsubsection{Development Challenges}
The use of serverless computing at the network's edge may significantly enhance resource utilization and result in cost savings. It is difficult to develop a serverless edge computing architecture owing to the limited and heterogeneous nature of edge resources.  The following design considerations should be taken into account to guarantee the architecture's performance.
\begin{itemize}
    \item \textbf{Heterogeneity:} Edge infrastructure's computing resources are diverse and widespread. The architecture needs to be able to meet the deployment and abstraction requirements of heterogeneous resources in edge networks to assure the on-demand supply of resources and the flexibility of serverless edge computing. These requirements include the protocol design between heterogeneous edge computing nodes, as well as the pooling and unified orchestration of heterogeneous computing resources.
    \item \textbf{Scalability:} Deploying a serverless framework on the edge requires to verify that its inherent scalability is not compromised. This is in contrast to the conventional serverless deployment that is carried out in the cloud. Further, since the number of mobile devices continues to rise exponentially, the architecture must be flexible enough to accommodate frequent replacement of edge computing nodes with serverless capabilities. When a new serverless-enabled edge computing node is deployed, the architecture is capable of recognizing and configuring the new node.
    \item \textbf{Performance:} Function prototypes are executed on virtual machines (VMs) or containers using the serverless computing architecture. This framework is too complicated to function properly in an edge context.
    Therefore, there is a possibility that it will not be able to properly ensure a low level of latency, particularly when the function is being invoked for the very first time. The performance of serverless edge computing architecture is to accommodate a high number of serverless services in resource-constrained edge contexts while assuring low response time.
    \item \textbf{Reliability:}  The reliability of computing that does not rely on servers is not very high. If a node fails, the requests that are associated with it may either be deleted or repeated several times. This does not lend itself well to supporting edge applications that need a high degree of reliability, such as the processing of medical data and the operation of autonomous vehicles. As a consequence of this, the architecture of serverless edge computing has to have high dependability and the ability to provide quick recovery in the event of a failure.
\end{itemize}
\subsubsection{Platforms \& Performance Metrics}
\begin{itemize}
\item \textbf{Resource Management:}
The resources are non-static and heterogeneous in a serverless framework. As the scheduling of resources and the distribution of those resources may be extremely difficult in fog computing, there is a need for further study in this area~\cite{mampage2022holistic-18}.
\begin{table*}[]
\caption{}
\label{tab: Conference and Journal}
\begin{tabular}{|l|l|l|}
\hline
Publisher Name                                                                                      & Type       & Number \\ \hline
IEEE International Conference on Fog Computing(ICFC)                                                & Conference & 2      \\ \hline
IEEE Global Communication  Conference                                                               & Conference & 1      \\ \hline
IEEE Network                                                                                        & Journal    & 1      \\ \hline
IEEE Transaction on Network and Service Management                                                  & Journal    & 1      \\ \hline
IEEE World Forum on Internet of Things (WF-IoT)                                                      & Conference & 1      \\ \hline
IEEE Internal Symposium on the World of Wireless, Mobile and Multimedia  Networks (WoWMoM)           & Conference & 1      \\ \hline
IEEE Internet Computing                                                                             & Journal    & 1      \\ \hline
IEEE  International Conference on Cloud Computing Technology and Science                            & Conference & 1      \\ \hline
IEEE World Congress on Services (SERVICES)                                                          & Conference & 1      \\ \hline
IEEE Annual Conference on Serverless Computing for Pervasive Cloud Edge Device Systems and Services & Conference & 1      \\ \hline
IEEE Conference and Exhibition on Global Telecommunications (GLOBECOM)                              & Conference & 1      \\ \hline
IEEE International Conference on Cloud Technology and Science (CloudCom)                            & Conference & 2      \\ \hline
IEEE Transactions on Parallel and Distributed Computing                                             & Journal    & 1      \\ \hline
IEEE Internet of Thing                                                                              & Journal    & 5      \\ \hline
IEEE International Conference on Edge Computing and Communications(EDGE)                            & Conference & 1      \\ \hline
IEEE Cloud Summit                                                                                   & Conference & 1      \\ \hline
IEEE International Conference on Fog Edge Computing(ICFEC)                                          & Conference & 1      \\ \hline
Internation Conference on Computing and Networking Technology                                       & Conference & 1      \\ \hline
IEEE INFOCOM                                                                                        & Conference & 1      \\ \hline
IEEE Internation Symposium on Cluster, Cloud and  Internet Computing (CCGrid)                       & Conference & 1      \\ \hline
IEEE Conference on Innovation in Cloud, Internet and Networks                                       & Conference & 1      \\ \hline
IEEE International Conference on Distributed Computing System                                       & Conference & 1      \\ \hline
International Workshop on Quality of Service                                                        & Workshop   & 1      \\ \hline
IEEE Journal of Biomedical and Health Informatics                                                   & Journal    & 1      \\ \hline
Internation Conference on Information Networking                                                    & Conference & 1      \\ \hline
IEEE Green Technologies Conference (IEEE-Green)                                                     & Conference & 1      \\ \hline
IEEE Annual Conference on Pervasive Computing and Communication Workshop (PerCom)                   & Conference & 1      \\ \hline
IEEE Transaction on Industrial Informatics                                                          & Journal    & 1      \\ \hline
Utility and Cloud Computing Campaign                                                                & Conference & 1      \\ \hline
ACM International Systems and Storage Conference                                                    & Conference & 1      \\ \hline
ACM International Conference on Parallel Processing Companion                                       & Conference & 1      \\ \hline
ACM International Conference on Internet of Thing Design and Implementation                         & Conference & 1      \\ \hline
ACM Internation Workshop on Edge Systems, Analytics, and Networking                                 & Workshop   & 1      \\ \hline
Computer Vision and Pattern Recognition                                                             & Journal    & 1      \\ \hline
USENIX Workshop                                                                                     & Workshop   & 1      \\ \hline
International Symposium on Cluster Computing and the Grid                                           & Conference & 1      \\ \hline
Future Generation Computer System                                                                   & Journal    & 2      \\ \hline
Pervasive and Mobile Computing                                                                      & Journal    & 1      \\ \hline
Journal of SuperComputing                                                                           & Journal    & 1      \\ \hline
Internet Technology Letetr                                                                          & Journal    & 1      \\ \hline
\end{tabular}
\end{table*}

 \item \textbf{Energy:}
Serverless computing holds the promise of increased resource efficiency via the use of fine-grained, embedded scaling in cloud environments. Cloud services are aware of the CPU and memory resources that have been subscribed to on the underlying serverless platform. However, when it comes to the edge, energy and latency are matters of worry, and even the most essential IoT use cases, such as Smart Farming, which are powered by battery-operated sensors and suffer from their short lifespan, are affected by these problems. Energy-aware function provisioning is required for Serverless at the edge; otherwise, the Serverless-enabled edge device may sacrifice energy to discover edge nodes with less supplied CPU and memory. When serverless computing moves to edge computing, it seems that cloud-driven performance metrics like scalability, concurrent requests, and startup time will need to be reexamined~\cite{mcgrath2017serverless-100}.
\item \textbf{Network Distribution:}
In serverless edge computing networks, it is not always possible to accomplish network growth by incorporating new resources. Therefore, effective network sharing mechanisms should be implemented to properly exploit the restricted and diverse serverless edge resources, which will ultimately result in lower operating costs and an improvement in network quality~\cite{antonopoulos2020bankruptcy-101},\cite{shafiei2019serverless-102}.
\item \textbf{Security:}
Serverless edge computing nodes, on the other hand, are not necessarily physically secure since they are vulnerable to direct assaults due to kernel vulnerabilities. This contrasts with cloud servers, which are placed in safe locations. In addition, functions are deployed in containers, which display less isolation than virtual machines (VMs), making them even more susceptible to attack.  It is possible to help with the introduction of identity and authentication processes and intrusion detection systems.
\end{itemize}
\subsubsection{Application Design}
\begin{itemize}
 \item \textbf{Application Service Management:} Research related to application management for serverless edge computing is important. Within the context of serverless edge computing, the service level, quality, and availability is the phenomena that are the most difficult to understand.
 \item \textbf{ Application Modeling:} It is difficult to simulate Serverless edge computing devices since apps have to acquire user data from a variety of IoT devices. As a result, there is a need for more study into the generation of generic applications.
\end{itemize}

\subsubsection{Emerging Trends and Open Issues}
\begin{itemize}
\item \textbf{Block-chain}
Serverless edge computing networks require a particular method to secure communication and increase data privacy since resource-constrained devices, and the inability to handle expensive security algorithms and firewalls are the two primary reasons for this necessity. The blockchain, which is a distributed ledger that is open and decentralized, is considered to be an exciting new technology. An unresolved question that warrants more investigation is how to incorporate blockchain technology into the serverless edge computing network.
\item \textbf{Edge Artificial Intelligence:}
Edge artificial intelligence (AI), which uses machine learning at the edge, is a rapidly growing topic for the development of future networks~\cite{chen2019deep-99}. It is possible to significantly cut operating expenses and streamline application development by using a serverless edge computing framework to carry out AI-related operations. However, this approach does come with certain complications, particularly in the form of cold start delays. AI functions are often I/O-bound due to the use of machine learning frameworks. I/O-bound functions, on the other hand, suffer lengthy cold start delays because it takes some time to load the machine learning library on containers. In situations like these, figuring out how to shorten the amount of time it takes for AI activities in serverless edge computing is a difficult challenge.
\item \textbf{No GPU Support}
The vast majority of commercial Serverless systems do not include application programming interfaces (APIs), which make it possible to offload computing to specialized hardware like GPUs and FPGAs, which may be more effective for the job at hand. Because of this limitation, it is impossible to use Serverless edge for GPU-based real-time streaming applications such as Smart Supply Chain Systems. These systems are outfitted with real-time information and monitor the entire supply chain, starting with the suppliers and ending with the retails~\cite{fraga2016review-103}. It would indicate that serverless architecture has to be rethought for such industries, particularly at the edge, by expanding the programming paradigms to be more comprehensive toward various hardware targets while maintaining the same degree of abstraction that we now have.
\end{itemize}

\section{Conclusion}\label{Conclusion}
In this paper, a systematic literature review is performed using current research articles based on serverless edge computing. An all-encompassing comprehension of the serverless edge computing frameworks as well as a discussion on open research issues, both of which may be accomplished with the help of this review. The SLR-based technique was implemented by running an exploratory query on the 192 published research papers. Finally, we have selected  47 research articles that are highlighting serverless edge computing. As per the current state of serverless edge computing, analysis is done on the basis of development, frameworks, metrics, and platforms for serverless edge computing. Moreover, a complete taxonomy has been presented in order to examine and compare the assessment criteria, trends analysis, and observations made based on chosen research. In addition, this study evaluated and summarised past review research in terms of development, performance measures, fog platform-based properties, communication mechanisms, and applications. Finally, we have identified possible future avenues and open research issues in the field of serverless edge computing. Table~\ref{tab: Conference and Journal} shows a list of Journals and Conferences.

\ifCLASSOPTIONcaptionsoff
  \newpage
\fi
\bibliographystyle{IEEEtran}
\bibliography{cas-refs.bib}

\begin{thebibliography}{10}
\providecommand{\url}[1]{#1}
\csname url@samestyle\endcsname
\providecommand{\newblock}{\relax}
\providecommand{\bibinfo}[2]{#2}
\providecommand{\BIBentrySTDinterwordspacing}{\spaceskip=0pt\relax}
\providecommand{\BIBentryALTinterwordstretchfactor}{4}
\providecommand{\BIBentryALTinterwordspacing}{\spaceskip=\fontdimen2\font plus
\BIBentryALTinterwordstretchfactor\fontdimen3\font minus \fontdimen4\font\relax}
\providecommand{\BIBforeignlanguage}[2]{{%
\expandafter\ifx\csname l@#1\endcsname\relax
\typeout{** WARNING: IEEEtran.bst: No hyphenation pattern has been}%
\typeout{** loaded for the language `#1'. Using the pattern for}%
\typeout{** the default language instead.}%
\else
\language=\csname l@#1\endcsname
\fi
#2}}
\providecommand{\BIBdecl}{\relax}
\BIBdecl

\bibitem{tang2020decentralized-1}
Q.~Tang, R.~Xie, F.~R. Yu, T.~Huang, and Y.~Liu, ``Decentralized computation offloading in iot fog computing system with energy harvesting: A dec-pomdp approach,'' \emph{IEEE Internet of Things Journal}, vol.~7, no.~6, pp. 4898--4911, 2020.

\bibitem{varghese2018next-2}
B.~Varghese and R.~Buyya, ``Next generation cloud computing: New trends and research directions,'' \emph{Future Generation Computer Systems}, vol.~79, pp. 849--861, 2018.

\bibitem{mao2017survey-3}
Y.~Mao, C.~You, J.~Zhang, K.~Huang, and K.~B. Letaief, ``A survey on mobile edge computing: The communication perspective,'' \emph{IEEE communications surveys \& tutorials}, vol.~19, no.~4, pp. 2322--2358, 2017.

\bibitem{chaudhry2020improved-4}
S.~R. Chaudhry, A.~Palade, A.~Kazmi, and S.~Clarke, ``Improved qos at the edge using serverless computing to deploy virtual network functions,'' \emph{IEEE Internet of Things Journal}, vol.~7, no.~10, pp. 10\,673--10\,683, 2020.

\bibitem{cicconetti2020decentralized-5}
C.~Cicconetti, M.~Conti, and A.~Passarella, ``A decentralized framework for serverless edge computing in the internet of things,'' \emph{IEEE Transactions on Network and Service Management}, vol.~18, no.~2, pp. 2166--2180, 2020.

\bibitem{ahmad2021container-7}
I.~Ahmad, M.~G. AlFailakawi, A.~AlMutawa, and L.~Alsalman, ``Container scheduling techniques: A survey and assessment,'' \emph{Journal of King Saud University-Computer and Information Sciences}, 2021.

\bibitem{aslanpour2021serverless-21}
M.~S. Aslanpour, A.~N. Toosi, C.~Cicconetti, B.~Javadi, P.~Sbarski, D.~Taibi, M.~Assuncao, S.~S. Gill, R.~Gaire, and S.~Dustdar, ``Serverless edge computing: vision and challenges,'' in \emph{2021 Australasian Computer Science Week Multiconference}, 2021, pp. 1--10.

\bibitem{glikson2017deviceless-10}
A.~Glikson, S.~Nastic, and S.~Dustdar, ``Deviceless edge computing: extending serverless computing to the edge of the network,'' in \emph{Proceedings of the 10th ACM International Systems and Storage Conference}, 2017, pp. 1--1.

\bibitem{kitchenham2004procedures-11}
B.~Kitchenham, ``Procedures for performing systematic reviews,'' \emph{Keele, UK, Keele University}, vol.~33, no. 2004, pp. 1--26, 2004.

\bibitem{dabbaghjamanesh2020real-12}
M.~Dabbaghjamanesh, A.~Moeini, A.~Kavousi-Fard, and A.~Jolfaei, ``Real-time monitoring and operation of microgrid using distributed cloud--fog architecture,'' \emph{Journal of parallel and distributed computing}, vol. 146, pp. 15--24, 2020.

\bibitem{gill2019transformative-13}
S.~S. Gill, S.~Tuli, M.~Xu, I.~Singh, K.~V. Singh, D.~Lindsay, S.~Tuli, D.~Smirnova, M.~Singh, U.~Jain \emph{et~al.}, ``Transformative effects of iot, blockchain and artificial intelligence on cloud computing: Evolution, vision, trends and open challenges,'' \emph{Internet of Things}, vol.~8, p. 100118, 2019.

\bibitem{tuli2020healthfog-14}
S.~Tuli, N.~Basumatary, S.~S. Gill, M.~Kahani, R.~C. Arya, G.~S. Wander, and R.~Buyya, ``Healthfog: An ensemble deep learning based smart healthcare system for automatic diagnosis of heart diseases in integrated iot and fog computing environments,'' \emph{Future Generation Computer Systems}, vol. 104, pp. 187--200, 2020.

\bibitem{xie2021serverless-15}
R.~Xie, Q.~Tang, S.~Qiao, H.~Zhu, F.~R. Yu, and T.~Huang, ``When serverless computing meets edge computing: architecture, challenges, and open issues,'' \emph{IEEE Wireless Communications}, vol.~28, no.~5, pp. 126--133, 2021.

\bibitem{mohanty2018evaluation-16}
S.~K. Mohanty, G.~Premsankar, M.~Di~Francesco \emph{et~al.}, ``An evaluation of open source serverless computing frameworks.'' \emph{CloudCom}, vol. 2018, pp. 115--120, 2018.

\bibitem{aslanpour2020performance-17}
M.~S. Aslanpour, S.~S. Gill, and A.~N. Toosi, ``Performance evaluation metrics for cloud, fog and edge computing: A review, taxonomy, benchmarks and standards for future research,'' \emph{Internet of Things}, vol.~12, p. 100273, 2020.

\bibitem{mampage2022holistic-18}
A.~Mampage, S.~Karunasekera, and R.~Buyya, ``A holistic view on resource management in serverless computing environments: Taxonomy and future directions,'' \emph{ACM Computing Surveys (CSUR)}, vol.~54, no. 11s, pp. 1--36, 2022.

\bibitem{jonas2019cloud-19}
E.~Jonas, J.~Schleier-Smith, V.~Sreekanti, C.-C. Tsai, A.~Khandelwal, Q.~Pu, V.~Shankar, J.~Carreira, K.~Krauth, N.~Yadwadkar \emph{et~al.}, ``Cloud programming simplified: A berkeley view on serverless computing,'' \emph{arXiv preprint arXiv:1902.03383}, 2019.

\bibitem{cassel2022serverless-20}
G.~A.~S. Cassel, V.~F. Rodrigues, R.~da~Rosa~Righi, M.~R. Bez, A.~C. Nepomuceno, and C.~A. da~Costa, ``Serverless computing for internet of things: A systematic literature review,'' \emph{Future Generation Computer Systems}, vol. 128, pp. 299--316, 2022.

\bibitem{ioini2020platforms-22}
N.~E. Ioini, D.~H{\"a}stbacka, C.~Pahl, and D.~Taibi, ``Platforms for serverless at the edge: a review,'' in \emph{European Conference on Service-Oriented and Cloud Computing}.\hskip 1em plus 0.5em minus 0.4em\relax Springer, 2020, pp. 29--40.

\bibitem{taibi2020patterns-24}
D.~Taibi, N.~El~Ioini, C.~Pahl, and J.~R.~S. Niederkofler, ``Patterns for serverless functions (function-as-a-service): A multivocal literature review,'' 2020.

\bibitem{eismann2020review-25}
S.~Eismann, J.~Scheuner, E.~Van~Eyk, M.~Schwinger, J.~Grohmann, N.~Herbst, C.~L. Abad, and A.~Iosup, ``A review of serverless use cases and their characteristics,'' \emph{arXiv preprint arXiv:2008.11110}, 2020.

\bibitem{yussupov2019systematic-26}
V.~Yussupov, U.~Breitenb{\"u}cher, F.~Leymann, and M.~Wurster, ``A systematic mapping study on engineering function-as-a-service platforms and tools,'' in \emph{Proceedings of the 12th IEEE/ACM International Conference on Utility and Cloud Computing}, 2019, pp. 229--240.

\bibitem{scheuner2020function-27}
J.~Scheuner and P.~Leitner, ``Function-as-a-service performance evaluation: A multivocal literature review,'' \emph{Journal of Systems and Software}, vol. 170, p. 110708, 2020.

\bibitem{li2022serverless-38}
Z.~Li, L.~Guo, J.~Cheng, Q.~Chen, B.~He, and M.~Guo, ``The serverless computing survey: A technical primer for design architecture,'' \emph{ACM Computing Surveys (CSUR)}, vol.~54, no. 10s, pp. 1--34, 2022.

\bibitem{kjorveziroski2021iot-48}
V.~Kjorveziroski, C.~Bernad~Canto, P.~Juan~Roig, K.~Gilly, A.~Mishev, V.~Trajkovik, and S.~Filiposka, ``Iot serverless computing at the edge: Open issues and research direction,'' \emph{Transactions on Networks and Communications}, 2021.

\bibitem{gadepalli2019challenges-49}
P.~K. Gadepalli, G.~Peach, L.~Cherkasova, R.~Aitken, and G.~Parmer, ``Challenges and opportunities for efficient serverless computing at the edge,'' in \emph{2019 38th Symposium on Reliable Distributed Systems (SRDS)}.\hskip 1em plus 0.5em minus 0.4em\relax IEEE, 2019, pp. 261--2615.

\bibitem{raith2023serverless}
P.~Raith, S.~Nastic, and S.~Dustdar, ``Serverless edge computing—where we are and what lies ahead,'' \emph{IEEE Internet Computing}, vol.~27, no.~3, pp. 50--64, 2023.

\bibitem{aslanpour2021serverless-8}
M.~S. Aslanpour, A.~N. Toosi, C.~Cicconetti, B.~Javadi, P.~Sbarski, D.~Taibi, M.~Assuncao, S.~S. Gill, R.~Gaire, and S.~Dustdar, ``Serverless edge computing: vision and challenges,'' in \emph{2021 Australasian Computer Science Week Multiconference}, 2021, pp. 1--10.

\bibitem{brereton2007lessons-28}
P.~Brereton, B.~A. Kitchenham, D.~Budgen, M.~Turner, and M.~Khalil, ``Lessons from applying the systematic literature review process within the software engineering domain,'' \emph{Journal of systems and software}, vol.~80, no.~4, pp. 571--583, 2007.

\bibitem{kitchenham2009systematic-29}
B.~Kitchenham, O.~P. Brereton, D.~Budgen, M.~Turner, J.~Bailey, and S.~Linkman, ``Systematic literature reviews in software engineering--a systematic literature review,'' \emph{Information and software technology}, vol.~51, no.~1, pp. 7--15, 2009.

\bibitem{kitchenham2013systematic-30}
B.~Kitchenham and P.~Brereton, ``A systematic review of systematic review process research in software engineering,'' \emph{Information and software technology}, vol.~55, no.~12, pp. 2049--2075, 2013.

\bibitem{stapic2012performing-31}
Z.~Stapic, E.~G. L{\'o}pez, A.~G. Cabot, L.~de~Marcos~Ortega, and V.~Strahonja, ``Performing systematic literature review in software engineering,'' in \emph{Central European Conference on Information and Intelligent Systems}.\hskip 1em plus 0.5em minus 0.4em\relax Faculty of Organization and Informatics Varazdin, 2012, p. 441.

\bibitem{kitchenham2012systematic-32}
B.~A. Kitchenham, ``Systematic review in software engineering: where we are and where we should be going,'' in \emph{Proceedings of the 2nd international workshop on Evidential assessment of software technologies}, 2012, pp. 1--2.

\bibitem{jatoth2015computational-33}
C.~Jatoth, G.~Gangadharan, and R.~Buyya, ``Computational intelligence based qos-aware web service composition: a systematic literature review,'' \emph{IEEE Transactions on Services Computing}, vol.~10, no.~3, pp. 475--492, 2015.

\bibitem{vakili2017comprehensive-34}
A.~Vakili and N.~J. Navimipour, ``Comprehensive and systematic review of the service composition mechanisms in the cloud environments,'' \emph{Journal of Network and Computer Applications}, vol.~81, pp. 24--36, 2017.

\bibitem{riaz2010experiences-35}
M.~Riaz, M.~Sulayman, N.~Salleh, and E.~Mendes, ``Experiences conducting systematic reviews from novices’ perspective,'' in \emph{14th International Conference on Evaluation and Assessment in Software Engineering (EASE)}, 2010, pp. 1--10.

\bibitem{batool2022software-36}
I.~Batool and T.~A. Khan, ``Software fault prediction using data mining, machine learning and deep learning techniques: A systematic literature review,'' \emph{Computers and Electrical Engineering}, vol. 100, p. 107886, 2022.

\bibitem{asghari2019internet-37}
P.~Asghari, A.~M. Rahmani, and H.~H.~S. Javadi, ``Internet of things applications: A systematic review,'' \emph{Computer Networks}, vol. 148, pp. 241--261, 2019.

\bibitem{cicconetti2018architectural-41}
C.~Cicconetti, M.~Conti, and A.~Passarella, ``An architectural framework for serverless edge computing: design and emulation tools,'' in \emph{2018 IEEE international conference on cloud computing technology and science (CloudCom)}.\hskip 1em plus 0.5em minus 0.4em\relax IEEE, 2018, pp. 48--55.

\bibitem{palade2019evaluation-42}
A.~Palade, A.~Kazmi, and S.~Clarke, ``An evaluation of open source serverless computing frameworks support at the edge,'' in \emph{2019 IEEE World Congress on Services (SERVICES)}, vol. 2642.\hskip 1em plus 0.5em minus 0.4em\relax IEEE, 2019, pp. 206--211.

\bibitem{das2018edgebench-57}
A.~Das, S.~Patterson, and M.~Wittie, ``Edgebench: Benchmarking edge computing platforms,'' in \emph{2018 IEEE/ACM International Conference on Utility and Cloud Computing Companion (UCC Companion)}.\hskip 1em plus 0.5em minus 0.4em\relax IEEE, 2018, pp. 175--180.

\bibitem{rajput2022edgefaasbench-74}
K.~R. Rajput, C.~D. Kulkarni, B.~Cho, W.~Wang, and I.~K. Kim, ``Edgefaasbench: Benchmarking edge devices using serverless computing,'' in \emph{2022 IEEE International Conference on Edge Computing and Communications (EDGE)}.\hskip 1em plus 0.5em minus 0.4em\relax IEEE, 2022, pp. 93--103.

\bibitem{ko2022function-47}
H.~Ko and S.~Pack, ``Function-aware resource management framework for serverless edge computing,'' \emph{IEEE Internet of Things Journal}, 2022.

\bibitem{hall2019execution-68}
A.~Hall and U.~Ramachandran, ``An execution model for serverless functions at the edge,'' in \emph{Proceedings of the International Conference on Internet of Things Design and Implementation}, 2019, pp. 225--236.

\bibitem{yadav2021browser-79}
D.~Yadav, S.~Maniar, K.~Sukhani, and K.~Devadkar, ``In-browser attendance system using face recognition and serverless edge computing,'' in \emph{2021 12th International Conference on Computing Communication and Networking Technologies (ICCCNT)}.\hskip 1em plus 0.5em minus 0.4em\relax IEEE, 2021, pp. 01--06.

\bibitem{wang2021wearmask-45}
Z.~Wang, P.~Wang, P.~C. Louis, L.~E. Wheless, and Y.~Huo, ``Wearmask: Fast in-browser face mask detection with serverless edge computing for covid-19,'' \emph{arXiv preprint arXiv:2101.00784}, 2021.

\bibitem{mendki2020evaluating-66}
P.~Mendki, ``Evaluating webassembly enabled serverless approach for edge computing,'' in \emph{2020 IEEE Cloud Summit}.\hskip 1em plus 0.5em minus 0.4em\relax IEEE, 2020, pp. 161--166.

\bibitem{baresi2019towards-9}
L.~Baresi and D.~F. Mendon{\c{c}}a, ``Towards a serverless platform for edge computing,'' in \emph{2019 IEEE International Conference on Fog Computing (ICFC)}.\hskip 1em plus 0.5em minus 0.4em\relax IEEE, 2019, pp. 1--10.

\bibitem{lin2021privacy-52}
S.-C. Lin and C.-H. Lin, ``Privacy-preserving serverless edge learning with decentralized small data,'' \emph{arXiv preprint arXiv:2111.14955}, 2021.

\bibitem{pfandzelter2020tinyfaas-51}
T.~Pfandzelter and D.~Bermbach, ``tinyfaas: A lightweight faas platform for edge environments,'' in \emph{2020 IEEE International Conference on Fog Computing (ICFC)}.\hskip 1em plus 0.5em minus 0.4em\relax IEEE, 2020, pp. 17--24.

\bibitem{cicconetti2022faas-65}
C.~Cicconetti, M.~Conti, and A.~Passarella, ``Faas execution models for edge applications,'' \emph{Pervasive and Mobile Computing}, vol.~86, p. 101689, 2022.

\bibitem{nastic2017serverless-6}
S.~Nastic, T.~Rausch, O.~Scekic, S.~Dustdar, M.~Gusev, B.~Koteska, M.~Kostoska, B.~Jakimovski, S.~Ristov, and R.~Prodan, ``A serverless real-time data analytics platform for edge computing,'' \emph{IEEE Internet Computing}, vol.~21, no.~4, pp. 64--71, 2017.

\bibitem{deng2021dependent-55}
S.~Deng, H.~Zhao, Z.~Xiang, C.~Zhang, R.~Jiang, Y.~Li, J.~Yin, S.~Dustdar, and A.~Y. Zomaya, ``Dependent function embedding for distributed serverless edge computing,'' \emph{IEEE Transactions on Parallel and Distributed Systems}, vol.~33, no.~10, pp. 2346--2357, 2021.

\bibitem{aslanpour2022energy-54}
M.~S. Aslanpour, A.~N. Toosi, M.~A. Cheema, and R.~Gaire, ``Energy-aware resource scheduling for serverless edge computing,'' in \emph{2022 22nd IEEE International Symposium on Cluster, Cloud and Internet Computing (CCGrid)}.\hskip 1em plus 0.5em minus 0.4em\relax IEEE, 2022, pp. 190--199.

\bibitem{tang2022distributed-56}
Q.~Tang, R.~Xie, F.~R. Yu, T.~Chen, R.~Zhang, T.~Huang, and Y.~Liu, ``Distributed task scheduling in serverless edge computing networks for the internet of things: A learning approach,'' \emph{IEEE Internet of Things Journal}, vol.~9, no.~20, pp. 19\,634--19\,648, 2022.

\bibitem{lakhan2022restricted-58}
A.~Lakhan, M.~A. Mohammed, A.~N. Rashid, S.~Kadry, K.~H. Abdulkareem, J.~Nedoma, R.~Martinek, and I.~Razzak, ``Restricted boltzmann machine assisted secure serverless edge system for internet of medical things,'' \emph{IEEE Journal of Biomedical and Health Informatics}, 2022.

\bibitem{hussain2019serverless-59}
R.~F. Hussain, M.~A. Salehi, and O.~Semiari, ``Serverless edge computing for green oil and gas industry,'' in \emph{2019 IEEE Green Technologies Conference (GreenTech)}.\hskip 1em plus 0.5em minus 0.4em\relax IEEE, 2019, pp. 1--4.

\bibitem{carpio2021engineering-61}
F.~Carpio, M.~Michalke, and A.~Jukan, ``Engineering and experimentally benchmarking a serverless edge computing system,'' in \emph{2021 IEEE Global Communications Conference (GLOBECOM)}.\hskip 1em plus 0.5em minus 0.4em\relax IEEE, 2021, pp. 1--6.

\bibitem{xie2022workflow-62}
R.~Xie, D.~Gu, Q.~Tang, T.~Huang, and F.~R. Yu, ``Workflow scheduling in serverless edge computing for the industrial internet of things: A learning approach,'' \emph{IEEE Transactions on Industrial Informatics}, 2022.

\bibitem{rausch2021optimized-63}
T.~Rausch, A.~Rashed, and S.~Dustdar, ``Optimized container scheduling for data-intensive serverless edge computing,'' \emph{Future Generation Computer Systems}, vol. 114, pp. 259--271, 2021.

\bibitem{yao2023performance-64}
X.~Yao, N.~Chen, X.~Yuan, and P.~Ou, ``Performance optimization of serverless edge computing function offloading based on deep reinforcement learning,'' \emph{Future Generation Computer Systems}, vol. 139, pp. 74--86, 2023.

\bibitem{jeon2021deep-73}
H.~Jeon, S.~Shin, C.~Cho, and S.~Yoon, ``Deep reinforcement learning for qos-aware package caching in serverless edge computing,'' in \emph{2021 IEEE Global Communications Conference (GLOBECOM)}.\hskip 1em plus 0.5em minus 0.4em\relax IEEE, 2021, pp. 1--6.

\bibitem{aske2018supporting-67}
A.~Aske and X.~Zhao, ``Supporting multi-provider serverless computing on the edge,'' in \emph{Proceedings of the 47th International Conference on Parallel Processing Companion}, 2018, pp. 1--6.

\bibitem{coviello2022dataxe-71}
G.~Coviello, K.~Rao, B.~Debnath, O.~Po, and S.~Chakradhar, ``Dataxe: A system for application self-optimization in serverless edge computing environments,'' in \emph{2022 IEEE International Conference on Pervasive Computing and Communications Workshops and other Affiliated Events (PerCom Workshops)}.\hskip 1em plus 0.5em minus 0.4em\relax IEEE, 2022, pp. 699--705.

\bibitem{lyu2022towards-69}
X.~Lyu, L.~Cherkasova, R.~Aitken, G.~Parmer, and T.~Wood, ``Towards efficient processing of latency-sensitive serverless dags at the edge,'' in \emph{Proceedings of the 5th International Workshop on Edge Systems, Analytics and Networking}, 2022, pp. 49--54.

\bibitem{sethunath2021joint-70}
M.~Sethunath and Y.~Peng, ``A joint resource allocation and request dispatch scheme for performing serverless computing over edge and cloud,'' in \emph{2021 IEEE 7th World Forum on Internet of Things (WF-IoT)}.\hskip 1em plus 0.5em minus 0.4em\relax IEEE, 2021, pp. 575--580.

\bibitem{cicconetti2021preliminary-72}
C.~Cicconetti, L.~Lossi, E.~Mingozzi, and A.~Passarella, ``A preliminary evaluation of quic for mobile serverless edge applications,'' in \emph{2021 IEEE 22nd International Symposium on a World of Wireless, Mobile and Multimedia Networks (WoWMoM)}.\hskip 1em plus 0.5em minus 0.4em\relax IEEE, 2021, pp. 268--273.

\bibitem{mistry2020demonstrating-85}
C.~Mistry, B.~Stelea, V.~Kumar, and T.~Pasquier, ``Demonstrating the practicality of unikernels to build a serverless platform at the edge,'' in \emph{2020 IEEE International Conference on Cloud Computing Technology and Science (CloudCom)}.\hskip 1em plus 0.5em minus 0.4em\relax IEEE, 2020, pp. 25--32.

\bibitem{smith2022fado-76}
C.~P. Smith, A.~Jindal, M.~Chadha, M.~Gerndt, and S.~Benedict, ``Fado: Faas functions and data orchestrator for multiple serverless edge-cloud clusters,'' in \emph{2022 IEEE 6th International Conference on Fog and Edge Computing (ICFEC)}.\hskip 1em plus 0.5em minus 0.4em\relax IEEE, 2022, pp. 17--25.

\bibitem{rausch2019towards-77}
T.~Rausch, W.~Hummer, V.~Muthusamy, A.~Rashed, and S.~Dustdar, ``Towards a serverless platform for edge $\{$AI$\}$,'' in \emph{2nd USENIX Workshop on Hot Topics in Edge Computing (HotEdge 19)}, 2019.

\bibitem{gill2021quantum-80}
S.~S. Gill, ``Quantum and blockchain based serverless edge computing: A vision, model, new trends and future directions,'' \emph{Internet Technology Letters}, p. e275, 2021.

\bibitem{li2022kneescale-81}
X.~Li, P.~Kang, J.~Molone, W.~Wang, and P.~Lama, ``Kneescale: Efficient resource scaling for serverless computing at the edge,'' in \emph{2022 22nd IEEE International Symposium on Cluster, Cloud and Internet Computing (CCGrid)}.\hskip 1em plus 0.5em minus 0.4em\relax IEEE, 2022, pp. 180--189.

\bibitem{paraskevoulakou2021leveraging-82}
E.~Paraskevoulakou and D.~Kyriazis, ``Leveraging the serverless paradigm for realizing machine learning pipelines across the edge-cloud continuum,'' in \emph{2021 24th Conference on Innovation in Clouds, Internet and Networks and Workshops (ICIN)}.\hskip 1em plus 0.5em minus 0.4em\relax IEEE, 2021, pp. 110--117.

\bibitem{cicconetti2020decentralized-40}
C.~Cicconetti, M.~Conti, and A.~Passarella, ``A decentralized framework for serverless edge computing in the internet of things,'' \emph{IEEE Transactions on Network and Service Management}, vol.~18, no.~2, pp. 2166--2180, 2020.

\bibitem{baresi2019towards-60}
L.~Baresi and D.~F. Mendon{\c{c}}a, ``Towards a serverless platform for edge computing,'' in \emph{2019 IEEE International Conference on Fog Computing (ICFC)}.\hskip 1em plus 0.5em minus 0.4em\relax IEEE, 2019, pp. 1--10.

\bibitem{huang2022mobility-83}
Y.~Huang, Z.~Lin, T.~Yao, X.~Shang, L.~Cui, and J.~Z. Huang, ``Mobility-aware seamless virtual function migration in deviceless edge computing environments,'' in \emph{2022 IEEE 42nd International Conference on Distributed Computing Systems (ICDCS)}.\hskip 1em plus 0.5em minus 0.4em\relax IEEE, 2022, pp. 447--457.

\bibitem{li2022joint-88}
Y.~Li, D.~Zeng, L.~Gu, K.~Wang, and S.~Guo, ``On the joint optimization of function assignment and communication scheduling toward performance efficient serverless edge computing,'' in \emph{2022 IEEE/ACM 30th International Symposium on Quality of Service (IWQoS)}.\hskip 1em plus 0.5em minus 0.4em\relax IEEE, 2022, pp. 1--9.

\bibitem{pelle2020operating-91}
I.~Pelle, J.~Czentye, J.~D{\'o}ka, A.~Kern, B.~P. Ger{\H{o}}, and B.~Sonkoly, ``Operating latency sensitive applications on public serverless edge cloud platforms,'' \emph{IEEE Internet of Things Journal}, vol.~8, no.~10, pp. 7954--7972, 2020.

\bibitem{kjorveziroski2022kubernetes-92}
V.~Kjorveziroski and S.~Filiposka, ``Kubernetes distributions for the edge: serverless performance evaluation,'' \emph{The Journal of Supercomputing}, pp. 1--28, 2022.

\bibitem{bac2022serverless-93}
T.~P. Bac, M.~N. Tran, and Y.~Kim, ``Serverless computing approach for deploying machine learning applications in edge layer,'' in \emph{2022 International Conference on Information Networking (ICOIN)}.\hskip 1em plus 0.5em minus 0.4em\relax IEEE, 2022, pp. 396--401.

\bibitem{zhang2020stoic-95}
M.~Zhang, C.~Krintz, and R.~Wolski, ``Stoic: Serverless teleoperable hybrid cloud for machine learning applications on edge device,'' in \emph{2020 IEEE International Conference on Pervasive Computing and Communications Workshops (PerCom Workshops)}.\hskip 1em plus 0.5em minus 0.4em\relax IEEE, 2020, pp. 1--6.

\bibitem{tutuncuouglu2022online-90}
F.~T{\"u}t{\"u}nc{\"u}o{\u{g}}lu, S.~Jo{\v{s}}ilo, and G.~D{\'a}n, ``Online learning for rate-adaptive task offloading under latency constraints in serverless edge computing,'' \emph{IEEE/ACM Transactions on Networking}, 2022.

\bibitem{pan2022retention-92}
L.~Pan, L.~Wang, S.~Chen, and F.~Liu, ``Retention-aware container caching for serverless edge computing,'' \emph{Proc. of IEEE INFOCOM, IEEE}, 2022.

\bibitem{mcgrath2017serverless-100}
G.~McGrath and P.~R. Brenner, ``Serverless computing: Design, implementation, and performance,'' in \emph{2017 IEEE 37th International Conference on Distributed Computing Systems Workshops (ICDCSW)}.\hskip 1em plus 0.5em minus 0.4em\relax IEEE, 2017, pp. 405--410.

\bibitem{antonopoulos2020bankruptcy-101}
A.~Antonopoulos, ``Bankruptcy problem in network sharing: Fundamentals, applications and challenges,'' \emph{IEEE Wireless Communications}, vol.~27, no.~4, pp. 81--87, 2020.

\bibitem{shafiei2019serverless-102}
H.~Shafiei, A.~Khonsari, and P.~Mousavi, ``Serverless computing: A survey of opportunities, challenges, and applications,'' \emph{ACM Computing Surveys (CSUR)}, 2019.

\bibitem{chen2019deep-99}
J.~Chen and X.~Ran, ``Deep learning with edge computing: A review,'' \emph{Proceedings of the IEEE}, vol. 107, no.~8, pp. 1655--1674, 2019.

\bibitem{fraga2016review-103}
P.~Fraga-Lamas, T.~M. Fern{\'a}ndez-Caram{\'e}s, M.~Su{\'a}rez-Albela, L.~Castedo, and M.~Gonz{\'a}lez-L{\'o}pez, ``A review on internet of things for defense and public safety,'' \emph{Sensors}, vol.~16, no.~10, p. 1644, 2016.

\end{thebibliography}

\begin{IEEEbiography}[
    {\includegraphics[width=1in,height=1.25in,clip,keepaspectratio]{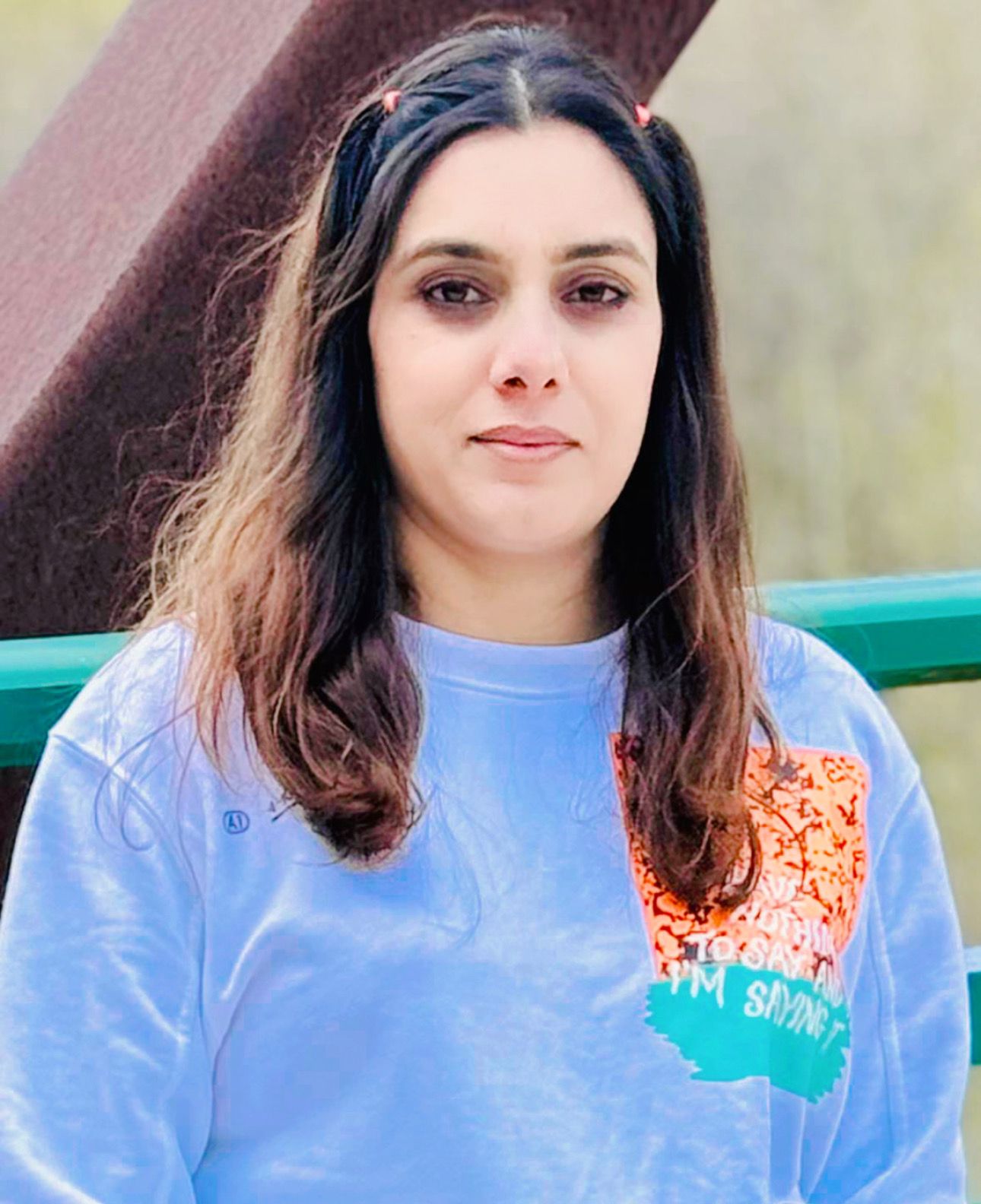}}]
    {Iqra Batool}
    Iqra Batool received the B.S. degree in Computer Science from the University of Azad Jammu and Kashmir and the M.S. degree in Software Engineering from Bahria University Islamabad, Pakistan. She is currently pursuing a Ph.D. in Computer Science at Western University, London, Ontario. Currently, Iqra Batool is a lecturer at Western University, where she teaches and conducts research in Wireless Communication, 5G/6G. Her research areas include serverless Edge Computing, Wireless Communication, and Software Fault Prediction.
\end{IEEEbiography}

\vspace{-500 pt}

\begin{IEEEbiographynophoto}{Sania kanwal}
  Sania Kanwal received a B.S. degree in Software Engineering from the University of Science and Technology Pakistan and an M.S. degree in Software Engineering from Bahria University Islamabad, Pakistan. Currently, she is pursuing a Ph.D. in Information Engineering at Southwest University of Science and Technology, Mianyang 621010, China. Her research areas include serverless edge computing and cyber attack detection in smart grids and IoTs.
\end{IEEEbiographynophoto}




\end{document}